\documentclass[letterpaper]{JHEP3}
\usepackage{cite}

\newcommand{\ket}[1]{|#1\rangle}
\newcommand{\CA}{{\cal A}}
\newcommand{\CB}{{\cal B}}
\newcommand{\CM}{{\cal M}}
\newcommand{\CN}{{\cal N}}
\newcommand{\CR}{{\cal R}}
\newcommand{\CT}{{\cal T}}
\newcommand{\CW}{{\cal W}}
\newcommand{\BR}{{\bf R}}
\newcommand{\BZ}{{\bf Z}}
\newcommand{\sla}{\mbox{\sl a}}
\newcommand{\slb}{\mbox{\sl b}}
\newcommand{\Tr}{{\rm Tr}}
\newcommand{\p}{{\partial}}
\title{Strings on $AdS_2$ and the High-Energy Limit\\ of Noncritical M-Theory}
\author{Petr Ho\v{r}ava and Cynthia A. Keeler\\
Berkeley Center for Theoretical Physics and Department of Physics\\
University of California, Berkeley, CA, 94720-7300\\
and\\
Theoretical Physics Group, Lawrence Berkeley National Laboratory\\
Berkeley, CA 94720-8162, USA}
\abstract{Noncritical M-theory in $2+1$ dimensions has been defined as a 
double-scaling limit of a nonrelativistic Fermi liquid on a flat 
two-dimensional plane.  Here we study this noncritical M-theory in the limit 
of high energies, analogous to the $\alpha'\to\infty$ limit of string 
theory.  In the related case of two-dimensional Type 0A strings, it has been 
argued that the conformal $\alpha'\to\infty$ limit leads to $AdS_2$ 
with a propagating fermion whose mass is set by the value of the RR flux.  
Here we provide evidence that in the high-energy limit, the natural ground 
state of noncritical M-theory similarly describes the $AdS_2\times S^1$ 
spacetime, with a massless propagating fermion.  We argue that the spacetime 
effective theory in this background is captured by a topological higher-spin 
extension of conformal Chern-Simons gravity in $2+1$ dimensions, consistently 
coupled to a massless Dirac field.  Intriguingly, the two-dimensional plane 
populated by the original nonrelativistic fermions is essentially the twistor 
space associated with the symmetry group of the $AdS_2\times S^1$ spacetime; 
thus, at least in the high-energy limit, noncritical M-theory can be 
nonperturbatively described as a ``Fermi liquid on twistor space.''}  

\begin{document}

\section{Introduction}

Noncritical string theories in $1+1$ dimensions (see 
\cite{Nakayama:2004vk,Ginsparg:1993is,Klebanov:1991qa,Alexandrov:2003ut,%
Martinec:2004td} for reviews) have long been a useful playground for studying 
stringy physics.  Unlike their ten-dimensional cousins, two-dimensional string 
theories are exactly solvable; thus, questions which are difficult to study in 
full string theory prove themselves more approachable in the theater of 
noncritical strings.  

In this paper, we use the setting of noncritical theories to examine 
some of the mysteries of M-theory.  In the full critical 
string case, we know very little about full M-theory beyond the
use of either nonperturbative dualities, or the low-energy limit as
described by eleven-dimensional supergravity.  Following the resurgence 
of interest in two-dimensional Type 0A and 0B string theories 
\cite{McGreevy:2003kb,Douglas:2003up,Takayanagi:2003sm}, in 
\cite{Horava:2005tt,Horava:2005wm} we proposed a nonperturbative 
definition of {\it noncritical M-theory\/} in $2+1$ dimensions, 
related to Type 0A and 0B strings in two dimensions by a string/M-theory 
duality.  The definition of noncritical M-theory as given in 
\cite{Horava:2005tt} is in terms of a double-scaling limit of a 
nonrelativistic Fermi liquid on a rigid two-dimensional plane.  In the 
double scaling limit, the number of fermions $N$ goes to infinity and 
the potential felt by the fermions becomes that of an an inverted harmonic 
oscillator.  Various ways of filling a Fermi sea correspond to various 
classical solutions of the theory.  This is also how two-dimensional Type 0A 
and 0B string vacua with a linear dilaton and RR flux are reproduced as 
solutions of noncritical M-theory: we recover their nonperturbative 
description as particular Fermi liquids of matrix-model eigenvalues.  In this 
correspondence, the role of the ``extra'' dimension of M-theory is played 
by the angular coordinate on the plane populated by the fermions: the Type 0A 
D0-brane charge is identified with the KK momentum along the extra dimension  
({\it \i.e.}, the angular momentum on the plane), 
mimicking the well-known correspondence from the critical case.%
\footnote{The full Type 0A theory in two dimensions has two separate RR 
fluxes \cite{Douglas:2003up}.  Making both nonzero simultaneously requires the 
presence of a nonzero number of long strings \cite{Maldacena:2005he}.  Only 
one of the RR-fluxes -- identified with a D0-charge -- plays a role in our 
definition of noncritical M-theory.  Whether or not the noncritical M-theory 
framework can be extended to incorporate the long strings and both RR fluxes 
is an interesting open question.}  

In addition to the two-dimensional string vacua, noncritical M-theory 
also contains a natural ground state, which we term the $\ket{M}$ state 
\cite{Horava:2005tt}.  This state is the noncritical analog of the 
eleven-dimensional M-theory vacuum solution.  As was shown in 
\cite{Horava:2005tt,Horava:2005wm}, 
the $\ket{M}$ state exhibits many features expected of a $2+1$ dimensional 
spacetime solution.  However, its description in terms of an effective gravity 
theory in a dynamical $2+1$-dimensional spacetime remains unknown.  In the 
related case of two-dimensional strings, the relation between the physical 
spacetime and the space populated by the fermions is quite subtle 
\cite{Ginsparg:1993is}.  The time dimension is the same between the two 
pictures.  However, the spatial Liouville dimension $x$ of the linear dilaton 
background is related to the spatial eigenvalue dimension $\lambda$ by an 
intricate integral transform, which can be viewed as an early form of string 
duality.  In noncritical M-theory, the situation is worse: so far, only the 
fermionic description has 
been developed, and how it maps to a physical spacetime picture is not yet 
understood.  It is the purpose of this paper to remedy this situation, and 
provide further evidence that the ground state of noncritical M-theory does 
indeed correspond to spacetime physics in $2+1$ dimensions, with an effective 
gravity description.  We will address this problem in the conformal limit of 
the theory, where our analysis will be facilitated by the larger symmetries of 
the system.  

In string theory, the conformal limit 
\cite{Strominger:2003tm,Ho:2004qp,Aharony:2005hm}   
corresponds to sending $\alpha'\to\infty$, which can also be interpreted as a 
high-energy limit \cite{Gross:1987kz,Gross:1987ar,Witten:1988zd} 
(see also \cite{Amati:1987uf,Amati:1987wq,Amati:1988tn}). 
This fact provides another motivation for this paper:  taking the high-energy 
limit of a mysterious theory in order to learn more about its underlying 
degrees of freedom is a classic strategy, pursued in string theory since its 
early days.  It has been widely speculated that in the high-energy limit, 
the theory might reveal an ``unbroken phase'' in which the massive string 
modes become massless \cite{Gross:1988ue,Witten:1988zd}.  Alternatively, one 
could probe the underlying degrees of freedom by heating the system to high 
temperature \cite{Atick:1988si}.  We applied this latter strategy to 
noncritical M-theory in \cite{Horava:2005wm}, 
and found a surprising connection between thermal noncritical M-theory and the 
topological strings of the A-model on the resolved conifold.  In this 
correspondence, the radius of the Euclidean time circle ({\it i.e.}, the 
inverse temperature) on the M-theory side plays the role of the A-model string 
coupling, a relation expected of topological M-theory 
\cite{Dijkgraaf:2004te}.  In the present paper, we complement the analysis of 
\cite{Horava:2005wm} and begin to probe the ground state of noncritical 
M-theory in another extreme regime, of high energies.  

This paper is organized as follows.  
After providing a quick review of noncritical M-theory in Section~\ref{non},  
and of the Type 0A conformal limit in Section~\ref{0A}, we will explore 
the same limit in the M-theory case  in Section~\ref{mlimit}, and argue that 
it describes an $AdS_2\times S^1$ spacetime.  The spectrum of propagating 
modes corresponds to the quanta of a single massless Dirac fermion on this 
background.  In Section~\ref{spacetime}, we address the question of an 
effective description of this system on the spacetime side.  First we embed 
the $AdS_2\times S^1$ as a vacuum solution to conformal $SO(3,2)$ Chern-Simons 
gravity in $2+1$ dimensions.  Then we extend the theory to a higher-spin 
Chern-Simons gauge theory, which not only incorporates the infinite symmetry 
of noncritical M-theory, but also allows a coupling to the propagating 
fermionic matter using the ``unfolded formalism'' of Vasiliev {\it et al.}
\cite{Vasiliev:1992gr,Vasiliev:1992ix,Shaynkman:2001ip} (see also 
\cite{Bekaert:2005vh} for a review).   Finally, we present our conclusions in 
Section~6, together with the amusing observation 
that from the spacetime point of view, the underlying Fermi liquid system 
can be viewed as living on twistor space associated with the conformal group 
$SO(3,2)$ of the $2+1$-dimensional dynamical spacetime.  

\section{Review of Noncritical M-Theory} \label{non}

\subsection{Definition as a Fermi Liquid}

Following \cite{Horava:2005tt}, we define noncritical M-theory by
starting with a regulated inverted harmonic oscillator potential on 
a two-dimensional plane $\BR^2$, with coordinates $\lambda_i$, $i=1,2$, 
filled with $N$ fermions.  The two-dimensional plane carries a fixed flat 
metric
\begin{equation}
ds^2=d\lambda_1^2+d\lambda_2^2.
\end{equation}
This metric is not dynamical.  The dynamical spacetime with a fluctuating 
metric field emerges as an effective structure associated with a particular 
solution of the theory.  This is exactly parallel to the case of 
two-dimensional string theory, wherein the eigenvalues of the matrix model 
live on a rigid space related to the spacetime Liouville dimension by an 
integral transform \cite{Ginsparg:1993is}.  

Then we take a double-scaling limit, simultaneously reducing the potential to 
an inverted harmonic oscillator while taking the number of fermions to 
infinity.  When the result of this process is written in the second-quantized 
language, with $\Psi (\lambda_i,t)$ a spinless fermion field, the appropriate 
action takes the nonrelativistic form 
\begin{equation}
S=\int dt d^2\lambda \left(i\Psi^\dag \frac{\partial\Psi}{\partial t}
-\frac{1}{2}\sum_{i=1,2}\frac{\partial\Psi^\dag}{\partial\lambda_i}
\frac{\partial\Psi}{\partial\lambda_i}
+\frac{1}{2}\omega_0^2\sum_{i=1,2}\lambda_i^2\Psi^\dag\Psi+\ldots\right),
\end{equation}
where $\omega_0$ is the fundamental frequency of the theory, and ``$\ldots$'' 
stand for nonuniversal regulating terms in the potential that are scaled away 
in the double-scaling limit.%
\footnote{The coordinates $\lambda_i$ before and after the double-scaling 
limit differ by an overall rescaling factor; in order to avoid notational 
clutter, we keep this factor implicit, and refer the reader to 
\cite{Horava:2005tt} for the exact technical details of the double-scaling 
limit.}
In the string-theory solutions of noncritical M-theory \cite{Horava:2005tt}, 
$\omega_0$ is related to $\alpha'$ by
\begin{equation}
\omega_0=\frac{1}{\sqrt{2\alpha'}}.
\end{equation}
In order to simplify our terminology, we will frequently refer to 
$1/(2\omega_0^2)$ as $\alpha'$ in the case of the M-theory vacuum as well.  

We will later also use the first quantized action, which is simply given by
\begin{equation}
\label{Maction} 
S=\frac{1}{2}\int dt
\sum_{i=1,2}\left(\dot{\lambda}_i^2+\omega_0^2\lambda_i^2\right).
\end{equation}

As explored in \cite{Horava:2005tt}, the richness of noncritical
M-theory comes from the freedom to pick any $N$ states to fill with
fermions.  However, there is still a most natural second quantized
ground state. We construct this state by filling every fermion with
individual energy below $-\mu$, while everything with higher energy
is kept empty.  This is the natural M-theory vacuum solution, and we will 
call it $\ket{M}$.  Its properties depend on the (double-scaled) value 
of the Fermi energy $\mu$, which plays the role of a coupling constant 
in the M-theory vacuum \cite{Horava:2005tt,Horava:2005wm}.

\subsection{Embedding of the Type 0A String}

In addition to the M-theory state, the vacua of two-dimensional 
Type 0A and 0B theories string theories are also solutions of noncritical 
M-theory as defined via the Fermi liquid system.  Here we will
concentrate on the embedding of the Type 0A linear dilaton vacuum with 
RR flux (see \cite{Horava:2005tt,Horava:2005wm} for 0B). In order to find the 
Type 0A 
state in noncritical M-theory, we first change variables from $\lambda_i$ to 
the polar coordinates $\lambda$ and $\theta$ (with $\lambda$ the radial 
coordinate).  Elementary separation of variables allows us to label the 
fermion quanta by their angular momentum $q$, and then discuss only the
dependence on the radial coordinate, $\lambda$. Thus, we are left
with a set of one-dimensional fermions, labelled by $q$, each living
in the following potential:
\begin{equation}
\label{potential}
 V(\lambda)=-\frac{1}{2}\omega_0^2\lambda^2
+\frac{M}{2\lambda^2},
\end{equation}
where $M=q^2-1/4$. The ground state describing the Type 0A vacuum with $q_0$ 
units of RR flux simply corresponds to placing all $N$ fermionic quanta in 
the lowest $N$ states with $q=q_0$ and taking the double scaling limit.  
(In the Type 0A language, the individual fermion corresponds to the 
open-string mode of a D0-$\overline{\rm D0}$ pair, and $q_0$ is the excess 
DO-brane charge.)  Since this solution of noncritical M-theory 
has been prepared such that all single-particle states with $q\neq q_0$ are 
kept empty in this ground state, all excitations in sectors with $q\neq q_0$ 
are infinitely energetic with respect to this ground state and therefore 
decouple, leaving precisely the excitations of the two-dimensional Type 0A 
vacuum at $q_0$ units of RR flux.  

This embedding of the Type 0A vacua as solutions in noncritical M-theory 
sheds new light on the M-theory ground state solution $\ket{M}$.  Indeed, 
we can think of $\ket{M}$ as a coherent sum of Type 0A vacua for all values 
of RR flux.  All single-particle (or hole) excitations are now finitely 
energetic with respect to the Fermi surface of $\ket{M}$, and represent 
physical excitations of the $\ket{M}$ state.  

This formal decomposition of the ground state of noncritical M-theory into 
Type 0A sectors is useful technically, for example in the evaluation of the 
vacuum energy of the solution \cite{Horava:2005tt,Horava:2005wm}.  It also 
leads to a crucial ``correspondence principle'': because the finite-energy 
excitations of every individual Type 0A sector carry finite energy in the 
$\ket{M}$ state, the effective 0A physics of each sector must be reproduced 
by the properties of the $\ket{M}$ state as well.  As one application of this 
correspondence principle, one can argue that since the vacua of Type 0A string 
theory can be described by an effective action containing two-dimensional 
gravity, the $\ket{M}$ state should have a gravitational description also.  
We will use this correspondence as one of our guiding principles throughout 
this paper.  

\subsection{Symmetries of the $\ket{M}$ and $\ket{0A}$ States}

Now that we have reviewed the basic noncritical M-theory setup, let
us discuss a few implications of its symmetries.  The theory has an underlying 
infinite-dimensional symmetry algebra $\CW$, first discussed in Section 8.2 
of \cite{Horava:2005tt}.  This algebra is the M-theory analog of the 
$w_\infty$ symmetry algebras known from two-dimensional string theories 
\cite{Ginsparg:1993is}.  $\CW$ arises from four basic conserved charges, 
given in the classical limit by
\begin{equation}
\sla_i=\frac{1}{\sqrt{2}}(p_i+\omega_0\lambda_i)e^{-\omega_0t},\qquad
\slb_i=\frac{1}{\sqrt{2}}(p_i-\omega_0\lambda_i)e^{\omega_0t}.
\end{equation}
Here $\lambda_i$ with $i=1,2$ are again the Cartesian coordinates on $\BR^2$, 
and $p_i$ are the conjugate momenta.  $a_i$ and $b_j$ satisfy commutation 
relations implied by the canonical Poisson brackets between the momenta 
$p_i$ and coordinates $\lambda_j$.  The four charges $(a_i,b_j)$ form a 
natural coordinate system on the phase space $\CT=\BR^4$ of the fermions.  

The full algebra $\CW$ has a basis consisting of the Weyl-ordered products 
of an arbitrary finite number of $\sla_i$ and $\slb_j$.  We can assign a 
``degree'' to the elements of this basis, simply defined as the total degree 
of the corresponding monomial in $\sla_i$ and $\slb_j$.  Linear combinations 
of the ten independent charges of degree two,  
\begin{equation}
\sla_1^2,\ \sla_2^2,\ \slb_1^2,\ \slb_2^2,\ \sla_1\sla_2,\ \slb_1\slb_2,\
\sla_1\slb_2,\ \sla_2\slb_1,\ \frac{1}{2}(\sla_1\slb_1+\slb_1\sla_1),\ 
\frac{1}{2}(\sla_2\slb_2+\slb_2\sla_2),
\end{equation}
form a finite-dimensional subalgebra in the full infinite symmetry algebra 
$\CW$.   This algebra of quadratic charges is isomorphic to the Lie 
algebra of $SO(3,2)$ or, equivalently, of the noncompact version $Sp(4,\BR)$ 
of the symplectic group.  Taking appropriate linear combinations of these 
charges, one can show that the full algebra $\CW$ (and, in particular, the 
$SO(3,2)$ subalgebra of quadratic charges) is maintained in the 
$\alpha'\to\infty$ limit.

Even though the Fermi liquid theory exhibits this large symmetry $\CW$, any  
given {\it solution\/} will typically break some of $\CW$.  In particular, 
those solutions that are described by a semiclassical Fermi surface will 
generally break $\CW$ to the subalgebra that preserves the Fermi surface.  
As an example, consider again the Type 0A string theory background with RR 
flux $q_0$.  For this solution of noncritical M-theory, the relevant 
symmetries in $\CW$ are those that commute with the angular momentum generator 
\begin{equation}\label{J}
J=\frac{1}{2\omega_0}\left(\sla_1\slb_2-\sla_2\slb_1\right)
\end{equation}
on the two-dimensional plane.  This is dictated by the fact that the Type 0A 
solution of M-theory corresponds to filling all available states of $J=q_0$ 
up to $\mu$, while keeping states with $J\neq q_0$ empty.  Out of the ten 
quadratic generators in $\CW$, four survive; $J$ itself, plus the three 
diagonal combinations
\begin{equation}
\sla_1^2+\sla_2^2,\qquad \slb_1^2+\slb_2^2,\qquad
\frac{1}{2}(\sla_1\slb_1+\slb_1\sla_1+\sla_2\slb_2+\slb_2\sla_2).
\end{equation}
The four quadratic charges that commute with $J$ form an 
$SL(2,\BR)\times U(1)$ subalgebra in the $SO(3,2)$ algebra of quadratic 
charges in $\CW$.  From the perspective of Type 0A string theory, the 
$SL(2,\BR)$ factor of the surviving symmetry algebra corresponds precisely to 
the generators of the ground ring.  Of course, the $SL(2,\BR)$ symmetry may 
be further broken by the level $\mu$ of the Fermi sea in the Type 0A vacuum.  
This embedding of the Type 0A symmetries into $\CW$ will be important below.

\section{Review of The Conformal Limit of the 0A Matrix Model}
\label{0A}

The conformal limit of two-dimensional Type 0A string vacua vith RR flux 
has been studied in \cite{Strominger:2003tm,Ho:2004qp,Aharony:2005hm}.  
Taking $\alpha'$ to infinity in the linear dilaton spacetime with RR flux $q$ 
leads to the $AdS_2$ geometry.  This limit can also be viewed as a 
near-horizon limit of an extremally charged two-dimensional black hole 
\cite{Gukov:2003yp,Davis:2004xb,Danielsson:2004xf}. 
In \cite{Aharony:2005hm}, Aharony and Patir provide further evidence for 
this behavior, by analyzing the spectrum of the model in this limit.   

By examining the potential in (\ref{potential}), we see that the limit of 
$\omega_0\to 0$ allows us to ignore the $\lambda^2$ term.  We can view this 
limit alternatively as probing small $\lambda$.  Either way, the quantum 
mechanics of the individual eigenvalues reduces in this limit to
\begin{equation}
\label{actcf}
S=\frac{1}{2}\int dt
\left(\dot{\lambda}^2-\frac{M}{\lambda^2}\right).
\end{equation}
This is a conformal field theory in 0+1 dimensions, studied a long time ago 
in \cite{deAlfaro:1976je}.  The ground state of the second-quantized Fermi 
liquid consists of $N$ eigenvalues occupying all available states up to 
Fermi energy $\mu=0$, which ensures that conformal invariance is maintained.  
(Conformal invariance would also result from completely emptying or completely 
filling the entire Fermi sea.)

Two results are of importance here:  first, the action (\ref{actcf}) is 
invariant under the conformal symmetry $SO(2,1)\sim SL(2,\BR)$, which of 
course is equivalent to the isometries of an $AdS_2$ spacetime.  The 
$SL(2,\BR)$ generators are
\begin{eqnarray}
H&=&\frac{1}{2}\left(\dot{\lambda}^2+\frac{M}{\lambda^2}\right)\nonumber\\
D&=&-\frac{1}{4}\left(\lambda\dot{\lambda}+\dot{\lambda}\lambda\right)+tH\\
K &=& \frac{1}{2}\lambda^2+2tD-t^2H.\nonumber
\end{eqnarray}
$H$ is the Hamiltonian following from the action (\ref{actcf}), and it of 
course exhibits a continuous spectrum.  The proposal of 
\cite{Strominger:2003tm} is to interpret this Hamiltonian on the $AdS_2$ dual, 
as the evolution operator in the Poincar\'e time.  The duality to $AdS_2$ 
suggests that we should consider the evolution with respect to the compact 
operator%
\footnote{In calling this generator $\widetilde H$, we differ slightly from 
the commonly accepted convention in the literature.}
\begin{equation} \label{R}
\widetilde H=\frac{1}{2}\left(\frac{1}{\CR}K+\CR H\right),
\end{equation}
where $\CR$ is an arbitrary constant scale, to be identified with the 
curvature radius of the $AdS_2$.  $\widetilde H$ represents the time evolution 
with respect to the {\it global time\/} on $AdS_2$
\cite{Strominger:2003tm,Ho:2004qp,Aharony:2005hm}.  $\widetilde H$ has a 
discrete spectrum, with the $n$-th level eigenvalue given by
\begin{equation} \label{Rspectrum}
h_n=\frac{1}{2}+n+\frac{|q|}{2}.
\end{equation}
On the $AdS_2$ side, this should be interpreted as the spectrum of propagating 
matter in global time.  Aharony and Patir have shown that this spectrum 
exactly matches the spectrum of a free Dirac fermion on $AdS_2$ with mass
\begin{equation}
\label{mss}
m=\frac{|q|}{2\CR}.
\end{equation}
In this sense, the conformal limit of the type $0A$ matrix model is dual to a 
theory on $AdS_2$ whose matter excitations are precisely those of a spinor 
field of mass given in (\ref{mss}).  In particular, the bosonic degrees of 
freedom of the tachyon do not survive in the conformal limit.  Indeed, from 
the point of view of the compact generator $\widetilde H$, the ground state 
of the system is empty, and there is no macroscopic Fermi sea and thus no 
collective bosonic modes.  In the near-horizon interpretation of this limit, 
this means that the propagating excitations of the Fermi sea do not make it 
to the near-horizon region of the black hole, leaving just the individual 
fermionic eigenvalues as the only propagating excitations in that regime.  

\section{Large $\alpha'$ Limit of Noncritical M-Theory}
\label{mlimit}

As we have discussed in Section~3, the ground state of noncritical M-theory 
can be viewed in polar coordinates as a coherent collection of ground states 
of Type 0A string theory, all filled to the common Fermi level $\mu$.  
We now take the conformal limit of this $\ket{M}$ state.  Thus, we set 
$\mu=0$ in order to maintain conformal invariance of the vacuum.  This leads 
to manifest $SL(2,\BR)$ invariance of the vacuum.  On the dual spacetime 
side, it is thus natural to expect that $AdS_2$ will be a part of the 
spacetime geometry.   The rest of the geometry can be inferred as follows.  
Invoking our ``correspondence principle,'' one can predict the spectrum of 
excitations of the $\ket{M}$ state in the conformal limit, from the knowledge 
of the Type 0A spectrum as reviewed in Section~4.  As $\alpha'\to\infty$, 
each individual Type 0A sector with fixed $q$ will contribute one copy of a 
massive fermion, with mass $m$ given by (\ref{mss}).  In the M-theory vacuum, 
we thus get an infinite collection of Dirac fermions on $AdS_2$, with masses 
\begin{equation}
\label{mssall}
m=\frac{|q|}{2\CR},\qquad q\in\BZ.
\end{equation}
This spectrum of masses represents a Kaluza-Klein tower, obtained from the 
reduction of a massess fermion in $2+1$ dimensions on $S^1$ of radius $2\CR$.  
Thus, we expect that noncritical M-theory in the conformal limit corresponds 
to spacetime that is $AdS_2\times S^1$.  The matching of the spectra of 
propagating modes will be one of the tests of this conjecture.  In the rest 
of the paper, we will subject this conjecture to several additional tests.  

\subsection{Symmetry Generators}

Let us begin our discussion of the large $\alpha'$, or small $\omega_0$, limit 
of noncritical M-theory by considering the symmetry generators on the fermion 
side.  As in the 0A case above, we find that the action (\ref{Maction}) 
simplifies in the small $\omega_0$ limit: 
\begin{equation}
\label{Mlimitaction} S=\frac{1}{2}\int dt
\left(\dot{\lambda}_1^2+\dot{\lambda}_2^2\right)
\end{equation}

Now, we would like to consider the $\ket{M}$ state with $\mu$ set 
exactly to zero.  If we think of the $\ket{M}$ state as a coherent collection 
of Type 0A ground states of all possible values of RR flux $q$, each of these 
Type 0A sectors will be filled up to Fermi energy $\mu=0$.  In the conformal 
limit, the entire $SL(2,\BR)$ respects the Type 0A vacuum, and the same will 
be true of the M-theory ground state.  In the $\alpha'\to\infty$ limit of 
noncritical M-theory, the generators of this $SL(2,\BR)$ subalgebra of $\CW$ 
are
\begin{eqnarray}
H&=&\frac{1}{4}(\sla_1+\slb_1)^2+\frac{1}{4}(\sla_2+\slb_2)^2
=\frac{1}{2}\left(\dot{\lambda}_1^2+\dot{\lambda}_2^2\right)\nonumber\\
K&=&\frac{1}{4\omega_0^2}(\sla_1-\slb_1)^2+\frac{1}{4\omega_0^2}(\sla_2
-\slb_2)^2
=\frac{1}{2}\left(\lambda_1-\dot{\lambda}_1t\right)^2+\frac{1}{2}
\left(\lambda_2-\dot{\lambda}_2t\right)^2\label{gensm}\\
D&=&\frac{1}{4\omega_0}\left(\sla_1^2+\sla_2^2-\slb_1^2-\slb_2^2\right)
=\frac{1}{4}\left(\lambda_1\dot{\lambda}_1+\dot{\lambda}_1\lambda_1
+\lambda_2\dot{\lambda}_2+\dot{\lambda}_2\lambda_2-2\dot{\lambda}_1^2t
-2\dot{\lambda}_2^2t\right).\nonumber
\end{eqnarray}
They all commute with the angular momentum generator $J$, 
\begin{equation}
\label{genj}
J=\frac{1}{2\omega_0}(\sla_1\slb_2-\sla_2\slb_1)=\frac{1}{2}\left(\lambda_1
\dot\lambda_2-\lambda_2\dot\lambda_1\right).
\end{equation}
It turns out that, from the point of view of the first-quantized formulation, 
these $SL(2,\BR)\times U(1)$ generators enjoy a special status among all 
quadratic charges:  they can be realized geometrically by a change of 
coordinates on the space of $(t,\lambda_1,\lambda_2)$ which preserves the 
foliation of this space by constant time slices.  The action 
(\ref{Mlimitaction}) is indeed symmetric under the following sets of 
transformations, generated by (\ref{gensm}) and (\ref{genj}): 
\begin{equation}
\begin{array}{rlll}
H:\quad&t'=t-\omega,\quad&\lambda_1'(t')=\lambda_1(t),\ \ 
&\lambda_2'(t')=\lambda_2(t)\\
D:\quad&t'=e^{-\omega}t,&\lambda_1'(t')=e^{-\omega/2}\lambda_1(t),
&\lambda_2'(t')=e^{-\omega/2}\lambda_2(t)\\
K:\quad&t'=\frac{t}{\omega t+1},&\lambda_1'(t')=(1+\omega t)^{-1}\lambda_1(t),
&\lambda_2'(t')=(1+\omega t)^{-1}\lambda_2(t) \\
J:\quad&t'=t,&\lambda_1'(t')=(\cos{\omega})\lambda_1-(\sin{\omega})
\lambda_2,\ \ 
&\lambda_2'(t')=(\cos{\omega})\lambda_2+(\sin{\omega})\lambda_1
\end{array}
\end{equation}
These four generators have one algebraic relation:
\begin{equation}
\frac{1}{2}(HK+KH)-D^2-J^2=\frac{1}{4}.
\end{equation}
The quadratic charges in $SO(3,2)$ that are not in this $SL(2,\BR)\times U(1)$ 
are not realized geometrically on $(t,\lambda_i)$.  This does not mean that 
they cannot survive as ``hidden'' symmetries, but we do not expect them to be 
realized by Killing symmetries of the gravitational background.  We shall see 
in Section~5 that this picture is indeed correct.  

As an example, let us consider what happens to a particular generator 
of $SO(3,2)$ that does not belong to the $SL(2,\BR)\times U(1)$ 
subalgebra:
\begin{equation}
H'=\dot{\lambda}_1^2-\dot{\lambda}_2^2.
\end{equation}
This generator acts on $(t,\lambda_i(t))$ via
\begin{equation}
H':\ t=t',\qquad\lambda_1'(t')=\lambda_1(t-\omega),\qquad\lambda_2'(t')
=\lambda_2(t+\omega).
\end{equation}
There is no coordinate transformation on the $\lambda_i$ and $t$
which can represent this symmetry.  It is a symmetry of the action,
but not one which can be represented geometrically.  One can easily check 
that the same holds true for the other generators of $SO(3,2)$ not in the 
$SL(2,\BR)\times U(1)$ subalgebra.  As such, we do not expect any of these 
generators to be associated with an isometry in the dual spacetime picture.  
$SL(2,\BR)\times U(1)$ of course matches the isometries of $AdS_2\times S^1$. 

\subsection{Spectrum Matching: Fermions on $AdS_2\times S^1$}

As in the Type 0A case, the spectrum of $H$ in (\ref{gensm}) is continuous, 
and corresponds to the Hamiltonian evolution in the Poincar\'e time on 
$AdS_2\times S^1$.  We again switch to the global time Hamiltonian $\widetilde 
H$, defined by (\ref{R}), now in terms of the M-theory operators $H$ and $K$ 
of (\ref{gensm}).  It is straightforward but reassuring to see that the 
spectrum of $\widetilde H$ matches that of a massless fermion on 
$AdS_2\times S^1$, as expected from our ``correspondence principle''.  
If we transform our action from (\ref{Mlimitaction}) via
\begin{equation}
\widetilde\lambda_i(\tau)=\frac{\sqrt{\CR}}{\sqrt{\CR^2+t^2}}\lambda(t),\qquad
\tau=\arctan{(t/\CR)},
\end{equation}
we find
\begin{equation}
S=\frac{1}{2}\int d\tau
\left[(\partial_\tau\widetilde\lambda_1)^2+(\partial_\tau\widetilde\lambda_2)^2
-\widetilde\lambda_1^2-\widetilde\lambda_2^2\right].
\end{equation}
Here $\tau$ corresponds to the global time, and the individual eigenvalues see 
a rightside-up planar harmonic oscillator potential.  $\widetilde H$ generates 
translations along $\tau$.  Its spectrum is 
\begin{equation}
h_{n,m}=\frac{1}{2}(n+m+1),
\end{equation}
where $n$ and $m$ both are non-negative integers. This is equivalent to the 
spectrum for the $1+1$ dimensional $\widetilde H$ as given in 
(\ref{Rspectrum}), if one allows $q$ to range over all integers.  
The second-quantized ground state of $\widetilde H$ with the rightside-up 
harmonic oscillator potential is again empty of all fermions, and there 
is no macroscopic Fermi surface with propagating bosonic excitations.  

Now, let us consider the spectrum of a free fermion, as calculated 
in global coordinates, on an $S^1$ fibered over $AdS_2$, as
suggested from the symmetry arguments above.  As shown in 
Appendix~\ref{appendix}, the only fibering which produces the same
spectrum is the direct product spacetime, that is $AdS_2\times S^1$.
Moreover, the spectrum matching requires the radius of the $S^1$ factor 
to equal $2\CR$, where $\CR$ is the curvature radius of $AdS_2$.

Thus, we conjecture that {\it in the $\alpha'\to\infty$ limit, the natural 
ground state $\ket{M}$ of noncritical M-theory describes a theory on a 
dynamical $AdS_2 \times S^1$ spacetime, with propagating matter described 
by a single free massless Dirac fermion.}

This conjecture leads to two remarkable phenomena: ({\it i}\/) We get a 
relativistic dual out of a nonrelativistic Fermi liquid,%
\footnote{Of course, this phenomenon is already present in two-dimensional 
string theory, in the duality between its Fermi liquid and spacetime 
descriptions, and therefore does not represent a novelty of noncritical 
M-theory.  What is perhaps new is that such a duality extends to a dimension  
higher than $1+1$.}
and ({\it ii}\/) the angular dimension on the flat plane populated by the 
nonrelativistic fermions corresponds under this duality to a fixed-radius 
circle fibered trivially over the $AdS_2$ base spacetime of string theory;  
this of course is the traditional behavior of the extra dimension in the 
simplest forms of string/M-theory duality.  

\subsection{A Family of Solutions with $SL(2,\BR)$ Symmetry}

In passing, we wish to point out that the conformal $AdS_2\times S^1$ vacuum 
belongs to an interesting multi-parameter family of solutions of 
noncritical M-theory, all of which share the $SL(2,\BR)$ symmetry of the 
Type 0A conformal vacua.  

Consider the ground state of the global-time Hamiltonian $\widetilde H$ of 
(\ref{R}) in the conformal limit of Type 0A theory.  As reviewed above, in 
the Fermi liquid picture this ground state is empty, {\it i.e.}, all the 
single-particle states are unoccupied by the fermions.  Another way of 
preparing a conformally invariant ground state would be to keep all 
single-particle states occupied, and treat the holes as elementary 
excitations.  Of course, in Type 0A theory these two solutions are isomorphic 
by the particle-hole duality, and we gain nothing by switching from one 
description to the other.  

In $2+1$ dimensions, the situation is more interesting.  The simplest ground 
state of the global-time Hamiltonian $\widetilde H$ is empty, leading to the 
$AdS_2\times S^1$ solution which is the main focus of the present 
paper.  Using particle-hole duality, it is again possible to switch between 
filled and empty states.  Doing so simultaneously for all values $q$ of the 
angular momentum would result in an equivalent solution.  However, 
unlike in $1+1$ dimensions, we now have the additional freedom of deciding 
separately for each value of $q$ whether all states are empty or full, without 
losing the $SL(2,\BR)$ symmetry of the state.  This leads to an interesting 
multi-parameter family of solutions, parametrized as follows.  We start 
with the ground state of $\widetilde H$ with all states empty, choose 
a sequence 
\begin{equation}
q_1<\ldots <q_n,
\end{equation}
and prepare a new state such that all available states with angular momentum 
$q$ in the range $q_{2k+1}\leq q<q_{2k}$ for $k=0,1,\ldots$ are occupied while 
those in sectors with $q_{2k}\leq q<q_{2k+1}$ remain empty.  

These solutions have $n$ sheets of the Fermi surface, located at $J=q_i$, 
$i=1,\ldots n$.  All of them inherit the $SL(2,\BR)$ symmetry from the Type 0A 
decomposition.  It would be very interesting to investigate the spacetime 
interpretation of this family of solutions.  When the individual Fermi 
surfaces are far separated, the physics of their collective excitations is 
almost decoupled.  However, since there is one common $SL(2,\BR)$ symmetry 
shared by the $n$ sheets of the Fermi surface, we expect the solution to have 
the structure of a multi-sheeted version of $AdS_2$, sharing the same boundary 
with a common conformal dual description.  This structure is very reminiscent 
of the multi-sheeted $AdS_5$ with a common CFT dual, studied in 
\cite{Aharony:2006hz,Kiritsis:2006hy}.  

\section{Spacetime Effective Action in the $AdS_2\times S^1$ Background}
\label{spacetime}

An observer in the $AdS_2\times S^1$ spacetime is not likely to describe 
the system in terms of the two-dimensional nonrelativistic Fermi liquid on 
the $\BR^2$ flat plane.  Instead, the physics of excitations that are 
sufficiently close to the ground state should be encoded in a spacetime 
effective action.  This action should contain spacetime gravity, it should 
have $AdS_2\times S^1$ as a solution, and should reproduce the symmetries 
and spectrum of the ground state.  It is the goal of this Section to 
propose a natural effective action that satisfies such constraints.  

We are working in a coordinate system $(x^\mu$), $\mu=0,1,2$, given by global 
coordinates $(x^0,x^1)=(t,\rho)$ on $AdS_2$, and a periodic coordinate $x^2=y$ 
of periodicity $2\pi$ on $S^1$.  The metric takes the following form, 
\begin{equation}
\label{adssc}
ds^2=-\CR^2\cosh^2\rho\,dt^2+\CR^2d\rho^2 +4\CR^2dy^2.
\end{equation}
The only nonzero component of the Einstein tensor in this spacetime is
\begin{equation}
R_{yy}-\frac{1}{2}Rg_{yy}=4.
\end{equation}
Thus, unlike an $AdS_3$ spacetime, $AdS_2\times S^1$ will not solve the vacuum 
Einstein equations for any value of the cosmological constant.  

With the hindsight afforded by AdS/CFT correspondence, it might be tempting 
to postulate the existence of a propagating $U(1)$ gauge field, whose flux 
through $AdS_2$ (and the dual flux through $S^1$) could provide just the 
right energy-momentum tensor to turn $AdS_2\times S^1$ into a solution of the 
coupled Einstein-Maxwell equations.%
\footnote{AdS/CFT correspondence for $AdS_n\times S^1$ spaces has been 
previously encountered in \cite{Klebanov:2004ya} in the connection with 
higher-dimensional noncritical superstrings.}
This is indeed possible, and leads to a solution of the coupled system
\begin{equation} \label{spaceequation}
R_{\mu\nu}-\frac{1}{2}Rg_{\mu\nu} = -\Lambda g_{\mu\nu} +8\pi
G_NT_{\mu\nu},
\end{equation}
where
\begin{equation}
T_{\mu\nu}=F_{\mu\lambda}F_\nu^\lambda-\frac{1}{4}g_{\mu\nu}
F_{\lambda\sigma}F^{\lambda\sigma}
\end{equation}
is the conventional energy-momentum tensor of the Maxwell Lagrangian.  

Let us consider $U(1)$ flux $F$ along the $AdS_2$ factor of the geometry, 
with $F$ proportional to the area two-form on $AdS_2$:
\begin{equation}
F_{t\rho}=-F_{\rho t}=f_0\cosh\rho
\end{equation}
The energy-momentum tensor is
\begin{equation}
T_{\mu\nu}=\frac{f_0^2}{2\CR^2}\left(\begin{array}{ccc}
\cosh^2\rho & 0 & 0\\
0 & -1 & 0 \\
0 & 0 & 1
\end{array}
\right).
\end{equation}
The Einstein equations are now satisfied by $AdS_2\times S^1$ if we pick
\begin{equation}
\label{lambf}
\Lambda = -\frac{4}{5\CR^2}, \qquad f_0=\frac{\CR}{\sqrt{5\pi G_N}}.
\end{equation}
Of course, the two form field strength $F$ is exact, with the gauge 
field given by
\begin{equation}
A= \frac{\CR\sinh\rho}{\sqrt{5\pi G_N}}dt; 
\end{equation}
the Maxwell equations for $A$ are trivially satisfied.  

The fact that there there is an electric flux through $AdS_2$ implies that 
the $S^1$ carries a dual, magnetic flux.  In $2+1$ dimensions, the dual to 
a $U(1)$ one-form gauge field $A$ is a scalar $\phi$, related to the 
field strength $F$ of $A$ by $\ast F=d\phi$.  By tracking this duality for 
our background (\ref{lambf}), one can easily see that $\phi$ has a flux 
through the $S^1$ factor of $AdS_2\times S^1$.  

However, even though $AdS_2\times S^1$ is a solution to the coupled 
Einstein-Maxwell system with negative cosmological constant in $2+1$ 
dimensions, this theory cannot be a good approximation to the effective theory 
describing the conformal limit of noncritical M-theory, for a simple reason.  
The analysis of the spectrum in Section~4 has revealed the existence of a 
single propagating matter field, a massless Dirac fermion in 
$AdS_2\times S^1$.  On the other hand, the spectrum of low-energy excitations 
of the Einstein-Maxwell theory would contain a propagating photon, an 
excitation of which there is no evidence on the noncritical M-theory side.  
Hence, we must look for another effective theory that has $AdS_2\times S^1$ as 
a solution, but with fewer propagating degrees of freedom.  

It turns out that the correct starting point is the Chern-Simons theory with 
$SO(3,2)$ gauge symmetry, {\it i.e.}, conformal gravity in $2+1$ dimensions.

\subsection{$AdS_2\times S^1$ in $SO(3,2)$ Chern-Simons Gravity}

In \cite{Horne:1988jf}, Horne and Witten extend the work of 
\cite{Witten:1988hc} and show that conformal gravity in $2+1$ dimensions can 
be rewritten as an $SO(3,2)$ Chern-Simons gauge theory of the conformal 
group.  We will now show that our $AdS_2\times S^1$ spacetime 
is a solution of this theory.  

Recall that conformal gravity on a $2+1$-dimensional manifold $\CM$ (with 
coordinates $x^\mu$) can be described by
\begin{equation}
S_{CS}=\frac{k}{4\pi}\int_\CM\Tr\left(A\wedge dA+\frac{2}{3}A\wedge A\wedge 
A\right),
\end{equation}
where $A$ is an $SO(3,2)$ Lie algebra-valued one-form gauge field.%
\footnote{Here ``Tr'' is the trace defined via the unique quadratic invariant 
on the simple group $SO(3,2)$.  The coupling $k$ is quantized because 
$\pi_3(SO(3,2))=\pi_3(SO(3))$ is nontrivial.  The precise quantization 
condition for $k$ will depend on the exact choice of the gauge group, 
{\it i.e.}, on whether we choose the $SO(3,2)$ group itself or one of its 
covers.  We shall briefly return to this point in Section~5.4 below.} 
We write $A$ in components as%
\footnote{We use essentially the same notation as \cite{Horne:1988jf}, 
with the only exception that we refer to the gauge fields associated to the 
special conformal transformations $\mathcal K_a$ as $\zeta_\mu^a$, while 
\cite{Horne:1988jf} used $\lambda_\mu^a$.}
\begin{equation}
\label{CSaction}
A_\mu=e_\mu{}^a\mathcal P_a+\omega_\mu{}^a\mathcal J_a+
\zeta_\mu{}^a\mathcal K_a + \phi_\mu\mathcal D. 
\end{equation}
Here $a$ runs over $0,1,2$ for each of $\mathcal P_a$, $\mathcal
J_a$, and $\mathcal K_a$.  In the interpretation of this theory as conformal 
gravity, $e_\mu{}^a$ are the components of the vielbein, while $\omega^a$ is 
the corresponding spin connection, Hodge-dualized in its internal Lorentz 
indices using the $\epsilon^{abc}$ tensor associated with 
$\eta_{ab}={\rm diag}(-1,1,1)$.  The commutation relations are 
\begin{equation}
\begin{array}{rlrl}
[\mathcal J_a,\mathcal J_b]&=\epsilon_{abc}\mathcal J^c, &
[\mathcal P_a,\mathcal P_b]&=[\mathcal K_a,\mathcal K_b]=
[\mathcal J_a,\mathcal D]=0, \cr
[\mathcal P_a,\mathcal K_b]&=\eta_{ab}\mathcal D-\epsilon_{abc}\mathcal J^c,
& & \cr
[\mathcal P_a,\mathcal J_b]& =\epsilon_{abc}\mathcal P^c, &
[\mathcal K_a,\mathcal J_b]&=\epsilon_{abc}\mathcal K^c,\cr
[\mathcal P_a,\mathcal D]& =\mathcal P_a,&
[\mathcal K_a,\mathcal D]&=-\mathcal K_a.
\end{array}
\end{equation}
The equations of motion of (\ref{CSaction}) are just the flatness conditions 
\begin{equation}\label{flatness}
F_{\mu\nu}=\partial_\mu A_\nu-\partial_\nu A_\mu+[A_\mu,A_\nu]=0.
\end{equation}

An interesting class of solutions to (\ref{flatness}) can be constructed as 
follows.  First we assume that our vielbein is invertible, and then we pick a 
gauge in which $\phi_\mu=0$.  In such circumstances, the equations of motion 
(\ref{flatness}) reduce to
\begin{eqnarray}
de^a-e_b\wedge\omega^{ab}&=&0 \label{Pflat}\\
d\zeta^a-\zeta_b\wedge \omega^{ab}&=&0 \label{Kflat}\\
e^a\wedge\zeta_a&=&0 \label{Dflat}\\
-d\omega^{ab}-\omega^{ac}\wedge\omega_c^{\
b}+e^a\wedge\zeta^b-e^b\wedge\zeta^a&=&0. \label{Jflat}
\end{eqnarray}
Before we discuss the embedding of our $AdS_2\times S^1$ spacetime into this
framework, let us discuss how $AdS_3$, $dS_3$, and the Minkowski space can 
be interpreted as solutions of this theory.  First, we note that 
Eqn.~(\ref{Pflat}) is simply the torsion-free condition. Setting $\zeta_\mu=
\zeta e_\mu$ where $\zeta$ is a constant, we find that Eqn.~(\ref{Kflat}) 
reduces to the torsion-free condition as well. Also, Eqn.~(\ref{Dflat}) is 
trivially satisfied.  Using $R^a_{\ b}=d\omega^a_{\ b}+\omega^a_{\ c}
\wedge\omega^c_{\ b}$, Eqn.~(\ref{Jflat}) becomes
\begin{equation}
\label{remain}
R^a_{\ b}-2\zeta e^a\wedge e_b=0.
\end{equation}
Consequently, when (\ref{remain}) is satisfied, the Einstein tensor reduces 
to
\begin{equation}
R_{\mu\nu}-\frac{1}{2}Rg_{\mu\nu}=-2\zeta g_{\mu\nu}.
\end{equation}
In this way, the vacuum with cosmological constant $\Lambda=2\zeta$ is 
embedded as a solution to conformal Chern-Simons gravity in $2+1$ 
dimensions.  This solution is described by the following gauge field, 
\begin{equation}
\label{allsolns}
A_\mu=e_\mu^{\ a}\left(\mathcal P_a
+\frac{\Lambda}{2}\mathcal K_a\right) +\omega_\mu^{\ a}\mathcal J_a.
\end{equation}
The explanation of the existence of such solutions is very simple. The 
solution (\ref{allsolns}) only excites gauge field components of a certain 
subalgebra of $SO(3,2)$.  When $\Lambda<0$, the nonzero components in 
(\ref{allsolns}) belong to $SO(2,2)\subset SO(3,2)$, for $\Lambda>0$ they span 
$SO(3,1)\subset SO(3,2)$, and if $\Lambda=0$ we get $ISO(2,1)\subset 
SO(3,2)$.  In all cases, the flatness of the $SO(3,2)$ connection reduces to 
the flatness in the corresponding subalgebra.  

Now, let us return to consider the embedding of our $AdS_2\times
S^1$ spacetime.  Our vielbein and spin connection components are
\begin{equation}
\begin{array}{rl}
e^0=&\CR\cosh \rho dt\\
e^1=&\CR\,d\rho\\
e^2=&2\CR\,dy \\
\omega^2=&-\sinh\rho\,dt.
\end{array}
\end{equation}
For the $\zeta^a$, we choose 
\begin{equation}
\zeta^0=-\frac{1}{2\CR^2}e^0,\qquad\zeta^1=-\frac{1}{2\CR^2}e^1,\qquad
\zeta^2=\frac{1}{2\CR^2}e^2.
\end{equation}
Thus, the $SO(3,2)$ gauge field that desribes $AdS_2\times S^1$ can finally be 
written as
\begin{equation}
A_\mu=e_\mu^{\ 0}\left(\mathcal P_0-\frac{1}{2\CR^2}\mathcal K_0\right)
+e_\mu{}^1\left(\mathcal P_1-\frac{1}{2\CR^2}\mathcal K_1\right)+e_\mu{}^2
\left(\mathcal
P_2+\frac{1}{2\CR^2}\mathcal K_2\right)+\omega_\mu{}^2 J_2.\label{gaugefield}
\end{equation}
Since we have again chosen $\phi_\mu=0$, the flatness conditions
$F_{\mu\nu}=0$ reduce to Eqns.~(\ref{Pflat}) -- (\ref{Jflat}).  
Simple algebra will show that these equations are satisfied.  
Thus, the background $A_\mu$ of (\ref{gaugefield}) is a solution of $SO(3,2)$ 
Chern-Simons gauge theory, and consequently of conformal gravity in $2+1$ 
dimensions.

The explanation for the existence of such a solution is again simple.  
(\ref{gaugefield}) corresponds to the embedding of $SL(2,\BR)\times U(1)$, the 
isometry of $AdS_2\times S^1$, into $SO(3,2)$.  In particular, the $U(1)$ 
factor is generated by 
\begin{equation}
\mathcal P_2+\frac{1}{2\CR^2}\mathcal K_2, 
\end{equation}
which indeed commutes with the three generators of $SL(2,\BR)$ also excited 
by the background gauge field (\ref{gaugefield}).  In more generality, for any 
embedding of a direct product $G_1\times G_2$ into the Chern-Simons gauge 
group $G$, the flatness conditions for $G$ factorize into the flatness in 
the $G_1$ and $G_2$ factors if the gauge field belongs to $G_1\times G_2$.  

The $SO(3,2)$ Chern-Simons gauge theory has the following good features, which 
make it a suitable starting point for constructing our effective action: 
({\it i}\/) $AdS_2\times S^1$ is a solution of this Chern-Simons theory, 
and the gauge group $SO(3,2)$ naturally coincides with the group of 
all the quadratic charges of the symmetry algebra $\CW$. ({\it ii}\/) The 
isometry $SL(2,\BR)\times U(1)$ of $AdS_2\times S^1$ is embedded in $SO(3,2)$ 
in precisely the manner expected from the Type 0A string theory and 
noncritical  M-theory arguments of Section~4. ({\it iii}\/) Unlike the 
Einstein-Maxwell system considered at the beginning of Section~5, the 
$SO(3,2)$ Chern-Simons gravity has no propagating bosonic degree of freedom,   
just as noncritical M-theory in the conformal limit.   

\subsection{Extension to Higher-Spin Gauge Theory}
\label{higherspin}

Despite its good properties, the $SO(3,2)$ Chern-Simons gauge theory cannot be 
the whole story, for two reasons: (1) $SO(3,2)$ is only a subalgebra of the  
infinite symmetry algebra of noncritical M-theory, and (2) we have seen 
evidence that in the conformal limit, the matter content of the noncritical 
M-theory vacuum is that of a propagating massless Dirac fermion.  In the 
conventional approach to Chern-Simons gravity, it is unknown how to couple 
second-quantized matter to the gravity sector as described by the gauge 
connection.  We shall now show that once the correct infinite symmetries of 
Problem~(1) are properly taken into account, Problem~(2) will acquire a 
natural solution as well.  

In order to resolve Problem~(1), we are in need of a Chern-Simons gravity 
theory based on an infinite-dimensional extension of $SO(3,2)$.  Remarkably, 
this theory is already available.  It is the bosonic version of a 
supersymmetric Chern-Simons theory of an infinite hierarchy of conformal 
higher-spin fields constructed in \cite{Shaynkman:2001ip}.  

Higher-spin gauge theories have a rich history, going back to the original 
work of Fradkin and Vasiliev \cite{Fradkin:1987ks} (see \cite{Bekaert:2005vh} 
for a review).  In $2+1$ dimensions, higher-spin gauge theories as 
Chern-Simons gauge theories were first written down by Blencowe 
\cite{Blencowe:1988gj}.  The higher-spin version of conformal Chern-Simons 
gravity in $2+1$ dimensions appeared first in the work of Pope and Townsend 
\cite{Pope:1989vj}, and Fradkin and Linetsky \cite{Fradkin:1989xt}.   We shall 
follow most closely the detailed construction by Shaynkman and Vasiliev 
\cite{Shaynkman:2001ip}; see also \cite{Didenko:2006zd}.%

The first thing we need is a convenient parametrization of the higher-spin 
symmetry algebra.  In order to construct this algebra, Shaynkman and Vasiliev 
\cite{Shaynkman:2001ip} first define operators%
\footnote{Despite appearances, $\hat a$ and $\hat a^+$ are not Hermitian 
conjugates of each other \cite{Shaynkman:2001ip}.}
$\hat a_{\alpha}$ and $\hat a^{+ \alpha}$, where the index $\alpha=1,2$ 
parametrizes the spinor representation of the Lorentz group in $2+1$ 
dimensions, and subject them to the commutation relations 
\begin{equation}
[\hat a_\alpha,\hat a^{+\beta}]=\delta_\alpha{}^\beta,\quad\quad[\hat a_\alpha,
\hat a_\beta]=[\hat a^{+\alpha},\hat a^{+\beta}]=0.
\end{equation}
Rather than use an operator realization, it is convenient to use techniques 
of noncommutative geometry on $\BR^4$ parametrized by commuting coordinates 
$a_{\alpha}$ and $a^{+\alpha}$ and endowed with the star product:
\begin{equation}\label{star}
(f \star g)(a,a^+)=f(a,a^+)\exp\left\{\frac{1}{2}\left(
\frac{\overleftarrow\p}{\p a_\alpha}\frac{\overrightarrow\p}{\p a^{+\alpha}}
-\frac{\overleftarrow\p}{\p a^{+\alpha}}\frac{\overrightarrow\p}{\p a_\alpha}
\right)\right\}g(a,a^+).
\end{equation}
This definition results in the following $\star$ commutators:
\begin{equation}
\label{starcom}
[a_\alpha,a^{+\beta}]_\star=a_\alpha\star a^{+\beta}-a^{+\beta}\star
a_\alpha=\delta_\alpha^\beta, \qquad
[a_\alpha,a_\beta]_\star=[a^{+\alpha},a^{+\beta}]_\star=0.
\end{equation}
The associative algebra of the $\star$-product defined by generators 
consisting of 
all powers of $a$ and $a^+$ is called $A_2$ in \cite{Shaynkman:2001ip}.  On 
$A_2$, one can define the structure of a Lie superalgebra, by first assigning 
even (or odd) grading to the generators given by even-degree (or odd-degree) 
monomials in $a_\alpha$ and $a^{+\alpha}$, and then defining the 
(anti)commutation relations via the $\star$-product (anti)commutator.   
In the rest of the paper, we shall call this higher-spin superalgebra 
$\widetilde\CW$.  

The subalgebra of quadratic charges in $\widetilde\CW$ is $Sp(4,\BR)$, which 
is isomorphic to the $SO(3,2)$ algebra.  It has a $\star$-product realization 
by
\begin{eqnarray*}
\mathcal P_a&=&\frac{1}{2}\sigma_a{}^{\alpha\beta}a_\alpha a_\beta, \qquad
\qquad\mathcal J_a=\frac{1}{2}\sigma_a{}^\alpha{}_\beta a_\alpha a^{+\beta},\\
\mathcal K_a&=&-\frac{1}{4}\sigma_{a\alpha\beta}a^{+\alpha}a^{+\beta},\ \qquad 
\mathcal D=\frac{1}{4}\left(a_\alpha a^{+\alpha}+a^{+\alpha}a_\alpha\right),
\end{eqnarray*}
where $\sigma_a{}^{\alpha\beta}$ are symmetric matrices given by 
\begin{equation}
\sigma_0{}^{\alpha\beta}=\pmatrix{1 &0\cr 0&1},\quad
\sigma_1{}^{\alpha\beta}=\pmatrix{0 &1\cr 1&0},\quad
\sigma_2{}^{\alpha\beta}=\pmatrix{1 &0\cr 0&-1},\quad
\end{equation}
and the spinor indices are raised and lowered as 
$c^\alpha=\epsilon^{\alpha\beta}c_\beta$, $c_\beta=\epsilon_{\alpha\beta}
c^\alpha$, with $\epsilon_{\alpha\beta}=-\epsilon_{\beta\alpha}$, 
$\epsilon_{12}=\epsilon^{12}=1$.

Clearly, this construction of $\widetilde\CW$ is very closely related to the 
construction of the $\CW$ symmetry algebra in noncritical M-theory which we 
reviewed briefly in Section~4.  More precisely, the infinite subalgebra 
$\CW_0$ of all even-degree charges in $\CW$ coincides with the maximal bosonic 
subalgebra in $\widetilde\CW$.  Our generators $\sla_i,b_j$ are related to 
$a_\alpha,a^{+\beta}$ of \cite{Shaynkman:2001ip} by a linear transformation 
that preserves the commutation relations, {\it i.e.}, by an $Sp(4,\BR)$ 
symplectomorphism of the phase space $\CT$ of the Fermi liquid.  

The conformal higher-spin theory that will be relevant for the conformal limit 
of noncritical M-theory is based on the bosonic higher-spin Lie algebra 
$\CW_0$.  This algebra contains generators that correspond to all integer 
spins; there is no evidence, in the noncritical M-theory vacuum that we 
consider here, of the half-integer fermionic spins.  We shall comment on a 
possible supersymmetric extension in Section~5.4.  

Thus, we consider the bosonic higher-spin theory, described again by the 
Chern-Simons action, 
\begin{equation}
\label{hcs}
S_{HCS}=\frac{k}{4\pi}\int\Tr\left(\CA\wedge d\CA+\frac{2}{3}\CA\wedge\CA
\wedge \CA\right).  
\end{equation}
with the gauge field one-form $\CA$ now taking values in the 
infinite-dimensional higher-spin Lie algebra $\CW_0$ of even-degree bosonic 
charges.%
\footnote{The ``$\Tr$'' in (\ref{hcs}) is defined as the bosonic restriction 
to $\CW_0$ of the natural supertrace defined on $\widetilde\CW$ 
(see \cite{Vasiliev:1986qx} and also Section~3 of \cite{Prokushkin:1999gc}).  
Conversely, one could try to keep the odd-degree generators as 
bosonic symmetries, {\it i.e.}, replace their anticommutation relations with 
commutation relations, as defined again via the $\star$-product algebra.  
However, the hypothetical gauge theory of $\CW$ would contain bosonic gauge 
field components of half-integer spins, leading to many conceptual 
difficulties; consequently, we will not consider this option in this paper.}
In general, $\CA$ can be expanded in components,
\begin{equation}
\label{expaa}
\CA(x|a,a^+)=\sum_{\ell=0}^\infty\sum_{m=0}^{2\ell}\frac{1}{m!(2\ell-m)!}
\CA_{\alpha_1\ldots\alpha_m}{}^{\alpha_{m+1}\ldots\alpha_{2\ell}}(x)
a^{+\alpha_1}\ldots a^{+\alpha_m}a_{\alpha_{m+1}}\ldots a_{\alpha_{2\ell}}.
\end{equation}
Each component $\CA_{\alpha_1\ldots\alpha_m}{}^{\alpha_{m+1}\ldots
\alpha_{2\ell}}(x)$ is a one-form on $\CM$.  

The equations of motion of (\ref{hcs}) yield the flatness condition,
\begin{equation} \label{oflat}
d\CA+\CA\star\wedge\CA=0.
\end{equation}
The gauge field $A$ of (\ref{gaugefield}), which describes the $AdS_2\times 
S^1$ solution of $SO(3,2)$ Chern-Simons gravity, can be embedded into the 
higher-spin theory by setting the components of $\CA$ in the $SO(3,2)$ 
subalgebra equal to $A$, and all others to zero.  In the $\star$-product 
language, our $AdS_2 \times S^1$ background is described by
\begin{equation}
\CA=\frac{1}{2}\left[e^a_\mu\sigma_a^{\alpha\beta}\left(a_\alpha
a_\beta + \frac{1}{4\CR^2}a_\alpha^+a_\beta^+\right)-e^2_\mu
\sigma_2^{\alpha\beta}(\frac{1}{2\CR^2}a^+_\alpha
a^+_\beta)+\omega_\mu^2\sigma_2{}^\alpha{}_\beta(a_\alpha
a^{+\beta})\right]dx^\mu.
\end{equation}
Thus, $AdS_2\times S^1$ is a solution of (\ref{oflat}).   

The theory (\ref{hcs}) is invariant under the full set of higher spin 
conformal transformations given by
\begin{equation} \label{higherspinxforms}
\delta \CA=d\varepsilon+[\CA,\varepsilon]_\star
\end{equation}
with $\varepsilon$ a scalar function with values in the infinite-dimensional 
higher-spin Lie algebra.  For any given solution $\CB$ of the equations 
of motion (\ref{oflat}), we are interested in its global symmetries, 
{\it i.e.}, gauge transformations $\varepsilon$ that preserve $\CB$:  
\begin{equation} 
\label{globsyms}
d\varepsilon+[\CB,\varepsilon]_\star=0.
\end{equation}
On a topologically trivial spacetime $\CM$, one global symmetry can be 
constructed for each element of the symmetry algebra, as follows.  Consider 
a solution $\CB$ of (\ref{oflat}).  Since $\CB$ is flat, it can be written as
\begin{equation}
\CB=g^{-1}(x)\star dg(x)
\end{equation}
for some function $g(x)$ with values in the Lie group $\CW_0$.  Then, for any 
fixed, constant element $\xi$ from the Lie algebra of $\CW_0$, 
\begin{equation}
\label{rigsymloc}
\varepsilon=g^{-1}(x)\star\xi\star g(x)
\end{equation}
is a symmetry of the background $\CB$, {\it i.e.}, a solution of 
(\ref{globsyms}).  

In the case of topologically nontrivial $\CM$, there could be obstructions 
against defining (\ref{rigsymloc}) globally over $\CM$.  Consequently, the 
actual symmetry can be reduced to a subalgebra.  As an example, consider for 
simplicity the Minkowski space, described as a solution of Chern-Simons theory 
in (\ref{allsolns}), and focus on the quadratic charges belonging to 
$SO(3,2)$.  Clearly, the spacetime-independent transformations 
\begin{equation}
\varepsilon(\mathcal P)=\xi^a\mathcal P_a
\end{equation}
solve (\ref{globsyms}) for any constant $\xi^a$; they represent the rigid 
translations of the Minkowski space.  Less trivially, one can show that 
\begin{eqnarray}
\varepsilon(\mathcal J)&=& \xi^a\left(\mathcal J_a+e_\mu{}^bx^\mu\epsilon_{abc}
\mathcal P^c\right),\nonumber\\
\varepsilon(\mathcal D)&=& \xi\left(\mathcal D-e_\mu{}^ax^\mu\mathcal P_a
\right),\label{glbsmm}\\
\varepsilon(\mathcal K)&=& \xi^a\left(\mathcal K_a-e_\mu{}^bx^\mu
\left(\epsilon_{abc}\mathcal J^c+\eta_{ab}\mathcal D\right)+\left(e_{\mu a}
e_\nu{}^bx^\mu x^\nu-\frac{1}{2}\delta_a^b x_\mu x^\mu\right)
\mathcal P_b\right)\nonumber
\end{eqnarray}
are also solutions of (\ref{globsyms}), and thus represent global symmetries 
of Minkowski space viewed as a solution of $SO(3,2)$ Chern-Simons gauge 
theory.%
\footnote{Note that only a smaller algebra corresponds to isometries of the 
background.}
Thus, on $\CM=\BR^3$, we see that the entire $SO(3,2)$ is a symmetry.  
However, if we compactify (say) $x^2$ on $S^1$, and require the 
global symmetries to be well-defined on $S^1$, only the linear combinations 
of (\ref{glbsmm}) that are independent of the $x^2$ coordinate survive the 
compactification.  The global symmetry group of the flat $\BR^2\times S^1$ 
background is reduced to $ISO(1,1)\times U(1)$.  

On $AdS_2\times S^1$ we are in a very similar situation.  Consider again 
the algebra of quadratic charges $SO(3,2)$.  If we sent the radius of $S^1$ 
to infinity, the global symmetry would correspond to the entire $SO(3,2)$.
The main difference compared to the Minkowski example is that on $AdS_2\times 
\BR$, the solutions to (\ref{globsyms}) are in fact periodic along the 
coordinate $x^2$ on $\BR$ with a fixed periodicity, set by the radius $\CR$ 
of $AdS_2$.  Which symmetries survive on $AdS_2\times S^1$ is thus 
determined by the radius of $S^1$ in units of $\CR$.  For a generic radius of 
$S^1$, $\CW_0$ is broken to the subalgebra of global charges that are 
independent of $x^2$.  

The quadratic charges that are independent of $x^2$ form the 
$SL(2,\BR)\times U(1)$ subalgebra of $SO(3,2)$.  It turns out that the 
quadratic charges that do depend on $x^2$ are periodic on $S^1$ whose radius 
is $\CR$ (or any integer multiple thereof).  In order to check this, we can 
go back to the coordinates $(t,\rho,y)$ of (\ref{adssc}), and find six 
solutions to (\ref{globsyms}) that depend on $y$, such as for example
\begin{eqnarray}
\varepsilon&=&\CR\,\sinh\rho\,\cos(2y)\left(\mathcal P_2-\frac{1}{2\CR^2}
\mathcal K_2\right)-\CR\,\cosh\rho\,\sin(2y)\left(\mathcal P_1+\frac{1}{2\CR^2}
\mathcal K_1\right)\nonumber\\
& &\qquad\qquad{}-\cosh\rho\,\cos (2y)\,\mathcal J_0+\sinh\rho\,\sin(2y)
\,\mathcal D,
\end{eqnarray}
Noting that $y$ is periodic with periodicity $2\pi$ when the $S^1$ radius is 
$2\CR$, we obtain the stated result.  Thus, we see that if the radius 
of $S^1$ is an integer multiple of $\CR$, all $SO(3,2)$ symmetries will be 
unbroken.  Since in noncritical M-theory the radius of the $S^1$ factor is 
twice the radius of $AdS_2$, the entire $SO(3,2)$ symmetry survives the 
compactification, in accord with our expectations from the Fermi liquid 
side discussed in Section~4.1.

In fact, there is an interesting refinement of the story.  The same 
$AdS_2\times S^1$ background would also be a solution of the supersymmetric 
extension of the theory, which would result from keeping both even- and 
odd-degree charges in the higher-spin algebra $\widetilde\CW$.  One can again 
ask what would be the periodicity of the odd-degree charges along $x^2$.  
It is intriguing that the odd-degree charges, and in particular the linear 
charges that correspond to the supercharges in the $OSp(1|4)$ supersymmetric 
extension of $SO(3,2)$, are all antiperiodic on $S^1$ of radius $\CR$.   This 
implies that the supercharges survive as global symmetries of  
$AdS_2\times S^1$ if the radius of $S^1$ is an {\it even\/} multiple of 
the $AdS_2$ radius.  We see that the radius of $S^1$ in the conformal limit 
of noncritical M-theory is precisely given by the minimal value for which 
the $AdS_2\times S^1$ background would be supersymmetric, if embedded into 
the supersymmetrized version of the higher-spin theory.  This suggests that 
our noncritical M-theory may be a simple $\BZ_2$ orbifold of a supersymmetric 
theory, on which we comment further in Section~5.4.  

\subsection{Coupling the Fermion}
\label{higherspinf}

Our analysis of the spectrum in Section~4 revealed that noncritical 
M-theory in the high-energy limit is not purely topological; the vacuum 
has at least one type of a propagating excitation, described by a 
second-quantized massless Dirac fermion field on the $AdS_2\times S^1$ 
background.  In the full effective action, this matter sector should couple 
consistently to the topological Chern-Simons sector of the theory.  The 
existence of such a coupling will represent another check of the proposed 
picture. 

The standard lore of topological gravity is that the system cannot be 
coupled to propagating matter.  This is sometimes avoided by representing 
matter in the first-quantized framework, essentially via a collection of 
Wilson lines coupled to the topological gauge field.  The difficulty 
essentially stems from the fact that in order to write down the equations 
of motion for a second-quantized propagating field, we must invert the 
vielbein; however, this is an unnatural procedure in Chern-Simons theory where 
we interpret the vielbein as a part of the Chern-Simons gauge field.  

Remarkably, this standard lore may no longer be valid once the gauge group 
becomes infinite dimensional.  Vasiliev {\it et al.} have shown 
\cite{Shaynkman:2001ip} (see \cite{Bekaert:2005vh} for a review) that 
propagating matter fields of low spins (in particular, a scalar and a spinor) 
can indeed be coupled the Chern-Simons higher spin gravity in $2+1$ 
dimensions.  

Again following \cite{Shaynkman:2001ip}, we first introduce the Fock vacuum 
$\ket{0}$ defined to satisfy $a_\alpha\ket{0}=0$.   In the $\star$-product 
realization, this vacuum can be described by a projector
\begin{equation}
|0\rangle\langle0|=4\exp{(-2a_\alpha a^{+\alpha})},
\end{equation}
which satisfies
\begin{equation}
a_\alpha\star|0\rangle\langle0|=|0\rangle\langle0|\star
a^{+\alpha}=0,\qquad |0\rangle\langle0|\star|0\rangle\langle0| =
|0\rangle\langle0|.
\end{equation}
In other words it is properly normalized, and annihilated on the
left by $a_\alpha$.  The full set of Fock states on this vacuum can be 
created by action on the left with $a^{+\alpha}$.  The matter fields on $\CM$ 
will be represented by a section of the Fock bundle over $\CM$, 
\begin{equation}
\label{matterf}
|\Phi(x|a^+)\rangle=\sum^\infty_{\ell=0}\frac{1}{\ell!}c_{\alpha_1...
\alpha_\ell}(x)a^{+\alpha_1}\ldots a^{+\alpha_\ell}\star |0\rangle\langle0|.
\end{equation}
For future reference, it will be natural to split $\ket{\Phi}$ into an even 
and odd part, 
\begin{eqnarray}
|\Phi_0\rangle&=&\sum^\infty_{n=0}\frac{1}{2n!}c_{\alpha_1...
\alpha_{2n}}(x)a^{+\alpha_1}\ldots a^{+\alpha_{2n}}\star |0\rangle\langle0|,
\nonumber\\
|\Phi_1\rangle&=&\sum^\infty_{n=0}\frac{1}{(2n+1)!}c_{\alpha_1...
\alpha_{2n+1}}(x)a^{+\alpha_1}\ldots a^{+\alpha_{2n+1}}
\star |0\rangle\langle0|.\label{splitt}
\end{eqnarray}
Each component $c_{\alpha_1\ldots\alpha_\ell}(x)$ is symmetric in all its 
indices.  Moreover, it is natural to consider the component fields 
$c_{\alpha_1\ldots\alpha_\ell}(x)$ as bosonic if $\ell$ is even and fermionic 
if $\ell$ is odd.  

The dynamics of the matter fields in a Chern-Simons background $\CA$ is 
encoded in the equations of motion, 
\begin{equation} \label{phiflat}
d|\Phi\rangle+\CA\star|\Phi\rangle=0.
\end{equation}
In terms of the components $c_{\alpha_1\ldots\alpha_\ell}$, these equations 
become
\begin{eqnarray}
2\partial_\mu c_{\alpha_1 \ldots
\alpha_\ell}&=&c_{\alpha_1\ldots\alpha_\ell\beta_1\beta_2}
e_\mu{}^{\beta_1\beta_2}
-\frac{\ell(\ell-1)}{4\CR^2}c_{\alpha_1\ldots\alpha_{\ell-2}}e_{\mu
\alpha_{\ell-1}\alpha_\ell}\nonumber\\
&&{}+\frac{\ell(\ell-1)}{2\CR^2}c_{\alpha_1\ldots\alpha_{\ell-2}}\sigma_{2\,
\alpha_{\ell-1}\alpha_\ell}e_\mu{}^2-(\ell+1)\omega_\mu{}^2\sigma_2{}^\beta
{}_{\alpha_1}c_{\alpha_2\ldots\alpha_\ell\beta},\label{compeq}
\end{eqnarray}
where $e_\mu{}^{\alpha\beta}=e_\mu^a\sigma_a{}^{\alpha\beta}$, and the 
symmetrization over $\alpha_1\ldots\alpha_\ell$ is kept implicit on the 
right-hand side of (\ref{compeq}).  Assuming that the vielbein is invertible, 
the infinite chain of equations (\ref{compeq}) can be used to solve 
algebraically for all the higher components $c_{\alpha_1\ldots\alpha_\ell}$ in 
terms of only two independend fields, given by the two lowest components $c$ 
and $c_\alpha$.  The full equation of motion (\ref{phiflat}) thus reduces to 
two dynamical equations for these remaining fields,
\begin{eqnarray}
\left(g^{\mu\nu}D_\mu D_\nu-\frac{1}{4\CR^2}\right)c(x)&=&0,\label{kgeq}\\
e^\mu_a\sigma^{a\beta}_{\phantom{a\beta}\alpha}\left(\partial_\mu
c_\beta+\omega_\mu^{\phantom{\mu}2}
\sigma_{2\phantom{\gamma}\beta}^{\phantom{2}\gamma}c_\gamma\right)&=&0.
\label{direq}
\end{eqnarray}
Here $g^{\mu\nu}$ is the inverse metric and $D_\mu$ is the covariant 
derivative, built from the spin connection and the vielbein.  

Some comments are now in order:
\begin{itemize}
\item
Eqns.~(\ref{kgeq}) and (\ref{direq}) are the Klein-Gordon and the Dirac 
equation for a massless spinor and scalar on $AdS_2\times S^1$.  All the 
components $c_{\alpha_1\ldots\alpha_\ell}$ with $\ell\geq 2$ are formed from 
derivatives of $c$ or $c_\alpha$, and thus do not constitute separate degrees 
of freedom themselves.  
\item
The equations of motion (\ref{compeq}) decouple components $c_{\alpha_1\ldots 
\alpha_\ell}$ with $\ell$ even from those with $\ell$ odd.  Thus, in our 
theory with $\CW_0$ gauge symmetry, the matter multiplet $\ket{\Phi}$ of 
(\ref{matterf}) is reducible, and can be split into two irreducible components 
given by $\ket{\Phi_0}$ and $\ket{\Phi_1}$ of (\ref{splitt}).  From the 
perspective of the subalgebra of quadratic charges $SO(3,2)\sim Sp(4,\BR)$, 
$\ket{\Phi_0}$ and $\ket{\Phi_1}$ correspond essentially to the two 
irreducible metaplectic representations of $Sp(4,\BR)$, sometimes called Di 
and Rac in the representation theory of this algebra.
\end{itemize}

In order to match our expected spectrum of noncritical M-theory in the 
conformal limit, as given in (\ref{Rspectrum}), we keep $\ket{\Phi_1}$ which 
contains the propagating massless Dirac fermion, but throw away $\ket{\Phi_0}$ 
which would contain the scalar.  

Second-quantized propagating matter can thus be coupled to Chern-Simons 
theory at the level of the equations of motion.  All matter interactions 
are mediated by the coupling to topological Chern-Simons theory, which itself 
does not propagate any physical degrees of freedom.  It is unclear, however, 
how to formulate an action principle for this system of equations of motion.  
Interestingly, Vasiliev {\it et al.} \cite{Prokushkin:1999gc} have suggested 
that such an action principle would have to be formulated not on $\CM$ but on 
a space that also includes $a_\alpha$.  

\subsection{Relation to a Supersymmetric Higher-Spin Theory}
\label{susy}

It is remarkable that the geometric properties of $AdS_2\times S^1$ that 
are required to match the Fermi liquid spectrum in the conformal limit, are 
precisely such that the radius of $S^1$ equals the minimum possible value 
compatible with unbroken supersymmetry of the solution.  This suggests that 
the effective theory that we propose for the description of the conformal 
limit of the noncritical M-theory vacuum is a simple $\BZ_2$ orbifold of a 
supersymmetric theory, with our $AdS_2\times S^1$ as a maximally 
supersymmetric solution.  

Such a supersymmetic extension of the effective theory of higher-spin 
Chern-Simons plus propagating massless matter can be easily constructed.  
In fact, it has already been written down by Shaynkman and Vasiliev in 
\cite{Shaynkman:2001ip}.%
\footnote{In \cite{Shaynkman:2001ip}, the focus is on $\CN=2$ supersymmetric 
theory; the $\CN=1$ version can be obtained by setting their $\hat k$ to zero.}
The gauge symmetry of this supersymmetric theory is given by the higher-spin 
conformal superalgebra $\widetilde\CW$.  The Chern-Simons gauge field 
$\widetilde\CA$ is now in the adjoint of $\widetilde\CW$.  It can be 
decomposed  into components, 
\begin{equation}
\widetilde\CA(x|a,a^+)=\sum_{\ell=0}^\infty\sum_{m=0}^{\ell}\frac{1}{m!
(\ell-m)!}\widetilde\CA_{\alpha_1\ldots\alpha_m}{}^{\alpha_{m+1}\ldots
\alpha_{\ell}}(x)a^{+\alpha_1}\ldots a^{+\alpha_m}a_{\alpha_{m+1}}\ldots 
a_{\alpha_{\ell}}.
\end{equation}
The component one-forms $\widetilde\CA_{\alpha_1\ldots
\alpha_m}{}^{\alpha_{m+1}\ldots\alpha_{\ell}}(x)$ are bosons if $\ell$ is 
even, and fermions for $\ell$ odd.  The lowest fermionic 
component of the gauge field corresponds to terms linear in $a_\alpha$ and 
$a^{+\alpha}$, and they describe the massless spin-$3/2$ gravitino of 
$\CN=1$ conformal supergravity in $2+1$ dimensions.  The full theory is then 
a higher-spin extension of $OSp(1|4,\BR)$ Chern-Simons supergravity.  

The Chern-Simons gauge sector is coupled to the massless matter 
supermultiplet, described by $\ket{\Phi}=\ket{\Phi_0}+\ket{\Phi_1}$ of 
Section~5.3.  As we have seen there, in the $AdS_2\times S^1$ background 
$\ket{\Phi_0}$ gives rise to a propagating boson and $\ket{\Phi_1}$ describes 
a propagating fermion.  Due to the periodicity properties of the generators 
of global symmetries, $AdS_2\times S^1$ is a supersymmetric solution of this 
theory if the radius of $S^1$ is an even multiple of the $AdS_2$ radius.  

On this $\CN=1$ supersymmetric theory, we can define the action of 
the orbifold group $\BZ_2=\{1,\Omega\}$ via
\begin{equation}
\Omega:\quad x^\mu\to x^\mu,\quad a_\alpha\to-a_\alpha,\quad 
a^{+\alpha}\to-a^{+\alpha},
\end{equation}
and extend it to the action on the fields by
\begin{eqnarray}
\Omega:\quad\CA(x|a,a^+)&\to&\CA(x|-a,-a^+),\\
\quad\ket{\Phi(x|a^+)}&\to&-\ket{\Phi(x|-a^+)}.
\end{eqnarray}
The orbifold projection by $\Omega$ projects out the odd-degree part of the 
gauge field $\widetilde\CA$ and the even-degree part $\ket{\Phi_0}$ of the 
matter multiplet.  Thus, our effective theory describing the conformal limit 
of the noncritical M-theory vacuum is an orbifold of the supersymmetric theory 
under this $\BZ_2$ action.  

This fact can perhaps be seen as another check of our proposal, for the 
following reason.  Just as in the critical spacetime dimension, 
two-dimensional Type 0A and 0B string theories are believed to be related to 
supersymmetric Type II cousins 
\cite{McGreevy:2003dn,Verlinde:2004gt,Takayanagi:2004ge,Seiberg:2005bx} by 
an orbifold construction.  Naturally, such supersymmetric two-dimensional 
theories could 
also be tied together into a supersymmetric version of noncritical M-theory, 
much like Type 0A and 0B backgrounds were in \cite{Horava:2005tt}.  If such a 
supersymmetric extension of noncritical M-theory exists, one would again 
expect it to be related to the nonsupersymmetric version by an orbifold.  It 
is intriguing that in the conformal limit, such a simple and natural extension 
does exist, at least at the level of the effective spacetime description, and 
that $AdS_2\times S^1$ extends naturally to a supersymmetric solution of it.%
\footnote{A $\BZ_2$-twisted version of the ground state of noncritical 
M-theory was studied in Section~9.4 of \cite{Horava:2005tt}, where it was 
shown that its vacuum energy vanishes  to all orders in the expansion in 
the coupling constant $1/\mu$.  Whether this feature is explained by some form 
of hidden supersymmetry in this state is not known.}
It would certainly be interesting to investigate these issues further. 

We conclude this Section with a few comments on the precise choice of the 
gauge group.  In particular, the quadratic charges in $\CW_0$ generate the 
Lie algebra of $SO(3,2)$, and the question is which of its covers should be 
chosen.  Conformal gravity in $2+1$ dimensions arises naturally as the gauge 
theory of $SO(3,2)$.  However, since our effective theory couples the gravity 
sector to a propagating fermion, the natural gauge group should be at least 
as large as ${\it Spin}(3,2)$, the double-cover of $SO(3,2)$.  The 
${\it Spin}(3,2)$ group is isomorphic to $Sp(4,\BR)$.  Since $Sp(4,\BR)$ also 
naturally arises on the Fermi-liquid side of noncritical M-theory, it 
might be tempting to propose it as the correct gauge group.  However, 
$Sp(4,\BR)$ itself has a nontrivial topological structure, with 
$\pi_1(Sp(4,\BR))=\BZ$.  As a result, it has a unique connected double cover, 
known as the ``metaplectic group'' ${\it Mp}(4,\BR)$.  The metaplectic group 
is {\it not\/} a matrix group; its smallest faithful representation, known as 
the ``metaplectic representation,'' is infinite-dimensional.  In fact, this 
metaplectic representation is equivalent to the Fock space representation of 
the $\hat a_\alpha$ and $\hat a^{+\alpha}$ operator algebra!  Thus, the 
metaplectic representation plays an essential role in the construction of our 
propagating matter multiplet.  Since it is only a projective representation 
of $Sp(4,\BR)$, it is natural to expect that the correct gauge group is not 
$Sp(4,\BR)$ but its double cover ${\it Mp}(4,\BR)$.  Perhaps, in noncritical 
M-theory in $2+1$ dimensions, ``M'' stands for ``metaplectic''!

\section{Conclusions}

In this paper, we have presented evidence suggesting that the ground state 
of noncritical M-theory, in the conformal limit, describes $AdS_2\times S^1$ 
spacetime.  The symmetries and spectrum of propagating modes in 
this limit are compatible with an effective theory given by the infinite 
higher-spin extension Chern-Simons gravity in $2+1$ dimensions, coupled to 
a propagating massless Dirac fermion.  This correspondence represents the 
simplest M-theory analog of the duality relation between the Liouville 
spacetime dimension and the eigenvalue coordinate as known from 
two-dimensional string theory.  

This correspondence leads to a nice matching of the ``extra dimension'' of 
noncritical M-theory in the $AdS_2\times S^1$ spacetime and in the Fermi 
liquid.  In the Fermi liquid picture, the ``extra dimension'' corresponds to 
the orbits of the $U(1)$ rotations of the rigid plane populated by the 
eigenvalues.  On the spacetime side, this $U(1)$ group becomes the group of 
translations along the $S^1$ factor of the $AdS_2\times S^1$ background, and 
the ``extra dimension'' acquires its traditional role as in critical 
string/M-theory.  In particular, the radius of the $S^1$ as measured by the 
spacetime metric is constant everywhere along $AdS_2$.  The fact that such a 
simple and intuitive picture emerges on the effective spacetime side lends 
further support to the original proposal of \cite{Horava:2005tt} that the 
extra dimension of M-theory indeed corresponds to the angular coordinate on 
the plane populated by the Fermi liquid.  

The conformal limit of the theory involves sending $\alpha'\to\infty$, and 
one can think of it as a certain high energy limit of the theory.  We have 
argued that in this limit, the effective spacetime theory is given by a 
higher-spin Chern-Simons gravity, coupled to a propagating fermionic 
degree of freedom.  It is intriguing that such a connection to higher-spin 
gauge theories emerges in the high-energy limit of noncritical M-theory.  
Indeed, the possibility of a close relation between higher-spin theories and 
the high-energy limit of critical string theory has been suspected for a long 
time (see, {\it e.g.}, \cite{Bekaert:2005vh,Sundborg:2000wp,Witten:2001js,%
Sezgin:2002rt,Klebanov:2002ja,Girardello:2002pp,Sagnotti:2003qa,%
Petkou:2004nu,Sagnotti:2005ns,Francia:2006hp} and references therein).  
We believe that the exactly solvable setting of noncritical M-theory in $2+1$ 
dimensions now provides an explicit testing ground for such ideas.%
\footnote{It also seems worth pointing out that our effective field theory 
for noncritical M-theory in $2+1$ dimensions is remarkably similar to the 
``holographic field theory'' proposal of \cite{Horava:1997dd}.  Indeed, the 
eleven-dimensional theory of \cite{Horava:1997dd} is a Chern-Simons gauge 
theory based on a higher-dimensional conformal superalgebra ${it OSp}(1|32)
\times {\it OSp}(1|32)$, coupled to propagating (fermionic) matter.  It thus 
appears that noncritical M-theory in $2+1$ dimensions might be a baby version 
of ``holographic field theory'' in the sense of \cite{Horava:1997dd}.}

Having found a dual interpretation of the ground-state solution of the Fermi 
liquid system in terms of a gravitational $AdS_2\times S^1$ background, 
one can turn the relation around, and ask the following question:  what is, 
from the perspective of an observer in $AdS_2\times S^1$, the interpretation 
of the dual space on which the Fermi liquid resides?  Since the double-scaling 
limit of the Fermi liquid involves taking a semiclassical limit, it is even 
more natural to look for an interpretation of the full phase space $\CT$, as 
parametrized either by $(\lambda_i,p_i)$ of Section~2 or equivalently by 
the conserved charges $(a_\alpha,a^{+\alpha})$ of Section~5.  Amusingly, it 
turns out that this phase space is precisely the twistor space associated 
with the $2+1$ dimensional conformal group $SO(3,2)$.  This relation can be 
made explicit by defining the twistor transform, which associates to a given 
element $(a_\alpha,a^{+\beta})$ of the twistor space a null vector $p^\mu$ at 
a spacetime point $x^\mu$, via
\begin{equation}
p^\mu=\sigma^{\mu\,\alpha\beta}a_\alpha a_\beta,\qquad 
a^{+\alpha}=x^\mu\sigma_{\mu}{}^{\alpha\beta}a_\beta.
\end{equation}
Perhaps the relation between the physical spacetime and the space of the Fermi 
liquid is simply related to such a twistor transform?  In any case, whether or 
not the twistor transform proves useful in this context, it is certainly 
true that the Fermi liquid lives on the twistor space associated with the 
conformal group $SO(3,2)$ of the $2+1$-dimensional spacetime.%
\footnote{This reference may need a bit of explanation.  In his after-dinner 
remarks at a Strings conference at (K)ITP Santa Barbara in the mid-1990's, 
Joe Polchinski proposed a ``Fermi liquid on twistor space'' as a half-joking 
answer to the question of ``what is string theory?''.}

This twistor perspective might shed some new light on another mysterious 
aspect of noncritical M-theory.  As shown in \cite{Horava:2005wm}, noncritical 
M-theory at finite temperature is closely related to the A-model topological 
strings on the resolved conifold, and plays essentially the role expected of 
topological M-theory.  In turn, topological M-theory has been conjecturally 
described by an effective gauge theory action in seven dimensions 
\cite{Dijkgraaf:2004te}.  It is puzzling why noncritical M-theory in $2+1$ 
dimensions should reproduce results expected of this seven-dimensional 
theory.  We do not have answers to this question, but we at least wish to 
point out one reason why seven dimensions should be relevant for noncritical 
M-theory.  As mentioned in Section~5.3, the equations of motion for the 
unfolded fermion (\ref{direq}) do not follow from any obvious action in three 
spacetime dimensions.  Attempts by Vasiliev {\it et al.} 
\cite{Prokushkin:1999gc} to resolve this problem suggest that an action might 
exist, but it would be naturally formulated on a bigger space that also 
includes the twistor coordinates.  In particular, one can view the 
Chern-Simons gauge field $\CA(x|a,a^+)$ of (\ref{expaa}) as living on the 
seven-dimensional space $\CT\times\CM$, the total space of the double 
fibration $\CT\leftarrow\CT\times\CM\rightarrow\CM$ familiar from twistor 
theory.

Various interesting open questions still remain.  In particular, it would be 
interesting to understand how to restore finite $\alpha'$ and study the full 
dynamics of the theory away from the conformal limit.  This should include 
the dynamics of other moduli, such as the ratio of the radii of 
$AdS_2\times S^1$.  
It would also be nice to understand what is the role, if any, of the Type 0A 
backgrounds with long strings and both values of the RR flux in the context 
of noncritical M-theory.  Answering this last question may require an 
understanding of the matrix model from which our Fermi liquid picture would 
follow.  Our effective description of the conformal limit of the theory in 
terms of $AdS_2\times S^1$ is a strong hint that a dual matrix model should 
exist, at least in this limit, in the form of a conformal quantum mechanics 
with a global $U(1)$ symmetry.  

\acknowledgments

We wish to thank Ofer Aharony, Eric Gimon, Tommy Levi, Oleg Lunin and Tassos 
Petkou for useful discussions.  This material is based upon work supported 
in part by NSF grants PHY-0244900 and PHY-0555662, DOE grant 
DE-AC03-76SF00098, an NSF Graduate Research Fellowship, and the Berkeley 
Center for Theoretical Physics.

\appendix
\section{Appendix A: Free Fermion Spectrum on $S^1$ Fibered over $AdS_2$} 
\label{appendix}

We wish to consider the spectrum of a free massless Dirac fermion on an 
$S^1$ fibration over $AdS_2$, assuming that the fibration preserves the 
$SL(2,\BR)$ symmetries of the base.  We work in global coordinates on 
$AdS_2$, and will study the spectrum with respect to the evolution in the 
global time on $AdS_2$.  The most general metric that preserves the 
$SL(2,\BR)$ symmetry of $AdS_2$ is
\begin{equation}
ds^2=-\CR^2\cosh^2\rho\,dt^2+\CR^2d\rho^2 +\CR^2(\gamma dy+\alpha\sinh\rho\,
dt)^2.
\end{equation}
The arbitrary constant $\alpha$ effectively measures the Kaluza-Klein flux 
of the off-diagonal metric components in the fibration.  
The other arbitrary constant $\gamma$ parameterizes the ratio of the radius 
of $S^1$ and the curvature radius of $AdS_2$ and $S^1$ components; we take 
the $S^1$ coordinate $y$ to run from 0 to $2\pi$.  Setting $\alpha=0$ would 
reproduce the direct product metric on $AdS_2\times S^1$.  Alternatively, the 
$AdS_3$ Hopf fibration would be obtained by setting $\alpha=\gamma=1$ and 
allowing $y$ to run over $\BR$.  

We will keep $\alpha$ and $\gamma$ arbitrary, in order to explore the full set 
of fibrations.  Additionally we will not impose boundary conditions yet to 
preserve generality.  The vielbein components 
\begin{eqnarray}
e^0&=&\CR\cosh \rho dt\\
e^1&=&\CR d\rho\\
e^2&=&\CR\gamma dy+\CR\alpha\sinh\rho dt
\end{eqnarray}
imply the spin connection components 
\begin{equation}
\begin{array}{l}
\omega_{01}=\frac{\alpha\gamma}{2}dy+\sinh\rho\left(\frac{\alpha^2}{2}
-1\right)dt\\
\omega_{02}=\frac{\alpha}{2}d\rho\\
\omega_{12}=-\frac{\alpha\cosh\rho}{2}dt.
\end{array}
\end{equation}
We will use the following $\gamma$ matrices, which have no explicit factors 
of $i$:
\begin{equation}
\gamma^0=i\sigma^2,\qquad\gamma^1=\sigma^1,\qquad\gamma^2=\sigma^3.
\end{equation}
Next, we need to calculate $\Gamma_\mu=\frac{1}{8}\omega_\mu{}^{ba}
\sigma^{ab}$:
\begin{equation}
\begin{array}{l}
\Gamma_y= -\frac{\alpha\gamma}{4}\sigma^3\\
\Gamma_\rho= \frac{\alpha}{4}\sigma^1\\
\Gamma_t=
\frac{1}{2}\left[(1-\frac{\alpha^2}{2})\right]\sinh\rho\sigma^3
-\frac{\alpha}{2}\cosh\rho i\sigma^2
\end{array}
\end{equation}
Finally, the massless Dirac equation
\begin{equation}
i\gamma^\mu\nabla_\mu\psi=(i\gamma^\mu\partial_\mu
+i\gamma^\mu\Gamma_\mu)\psi=0
\end{equation}
becomes
\begin{equation}
\left(\begin{array}{cc}
\cosh\rho\frac{1}{\gamma}\partial_y+\frac{\alpha}{4}\cosh\rho &
\partial_t+\cosh\rho\partial_\rho-\frac{\alpha}{\gamma}\sinh\rho\partial_y
-\frac{1}{2}\sinh\rho\\
-\partial_t+\cosh\rho\partial_\rho+\frac{\alpha}{\gamma}\sinh\rho\partial_y
-\frac{1}{2}\sinh\rho
&-\cosh\rho\frac{1}{\gamma}\partial_y+\frac{\alpha}{4}\cosh\rho
\end{array}\right)\psi=0.
\end{equation}
We will make the coordinate change $\cosh\rho=1/\cos\theta$
additionally stipulating $\sinh\rho=-\tan\theta$.  This gives
\begin{equation}
\left(\begin{array}{cc}
\sec\theta\frac{1}{\gamma}\partial_y+\frac{\alpha}{4}\sec\theta &
\partial_t-\partial_\theta
+\frac{\tan\theta}{2}+\frac{\alpha}{\gamma}\tan\theta\partial_y\\
-\partial_t-\partial_\theta
+\frac{\tan\theta}{2}-\frac{\alpha}{\gamma}\tan\theta\partial_y &
-\sec\theta\frac{1}{\gamma}\partial_y+\frac{\alpha}{4}\sec\theta
\end{array}
\right)\psi=0.
\end{equation}
Now, let us choose $\psi$ such that
\begin{equation}
\psi=e^{i\beta y}e^{-i\omega t}Y(\theta)\left(
\begin{array}{c}
e^{i\omega\theta}X_1(\theta)u(\theta)\\
e^{-i\omega\theta}X_2(\theta)v(\theta)
\end{array}
\right),
\end{equation}
where
\begin{equation}
\partial_\theta Y=\frac{1}{2}\tan\theta Y, \quad\partial_\theta
X_1=-i\frac{\alpha\beta}{\gamma}\tan\theta X_1, \quad\partial_\theta
X_2=i\frac{\alpha\beta}{\gamma}\tan\theta X_2.
\end{equation}
This choice of $\psi$ reduces our Dirac equation to
\begin{eqnarray}
\left(i\frac{\beta}{\gamma}+\frac{\alpha}{4}\right)\sec\theta 
e^{2i\omega\theta}\frac{X_1}{X_2}u-\partial_\theta v&=&0,\\
-\frac{X_1}{X_2}e^{2i\omega\theta}\partial_\theta u
+\left(-i\frac{\beta}{\gamma}+\frac{\alpha}{4}\right)\sec\theta v&=&0.
\end{eqnarray}
These coupled equations reduce to the single second-degree equation
\begin{equation}
\left(\frac{\alpha^2}{16}+\left(\frac{\beta}{\gamma}\right)^2\right)u
=\left(2i\omega\cos^2\theta-(1+2i\frac{\alpha\beta}{\gamma})\sin\theta
\cos\theta\right)\partial_\theta u + \cos^2\theta\partial_\theta^2u.
\end{equation}
Following \cite{Aharony:2005hm}, we change variables to
$z=(1+\tan\theta)/2$ which gives us
\begin{equation}\label{z}
\left(\frac{\alpha^2}{16}+\frac{\beta^2}{\gamma^2}\right)u+z(1-z)u''
+\left[\omega+\left(1-2i\alpha\frac{\beta}{\gamma}\right)\left(\frac{1}{2}
-z\right)\right]=0.
\end{equation}
Note that the $\alpha=0$ case reduces to Eqn.~(A.17) in 
\cite{Aharony:2005hm}, provided we set $\frac{\beta}{\gamma}=m\CR$. Although
our Dirac norm contains instead a factor of $1/\cos^3\theta$, we
still have the same requirement for $u$ to vanish at $\theta=\pm
\pi/2$, or $z\to\pm\infty$. Thus, for the particular case of $\alpha=0$, we 
can use the result of \cite{Aharony:2005hm} that
\begin{equation}
|\omega|=|\frac{\beta}{\gamma}|+\frac{1}{2}+n.
\end{equation}
Now, even in the $\alpha=0$ case, we must also consider the
restrictions the boundary conditions for $\psi$ in $y$ put on
$\beta$. Assuming periodicity of the fermions, we find
\begin{equation}
\beta=\frac{q}{2},
\end{equation}
which gives us exactly the spectrum (\ref{Rspectrum}) if we also set
$\gamma=2$.

We should also check that no other combination of $\alpha$ and
$\gamma$ produces the same spectrum.  Let us proceed by comparing
Equation (\ref{z}) to the generic form for a hypergeometric equation
\begin{equation}
z(1-z)u''+\left[c-(a+b+1)z\right]u'-abu=0.
\end{equation}
We can match Equation (\ref{z}) to this equation by choosing
\begin{equation}
a=-i\alpha\frac{\beta}{\gamma}-s,\qquad
b=-i\alpha\frac{\beta}{\gamma}+s,\qquad
c=\omega+\frac{1}{2}-i\alpha\frac{\beta}{\gamma}
\end{equation}
with
\begin{equation}
s=\sqrt{\frac{\alpha^2}{16}-\frac{\alpha^2\beta^2}{\gamma^2}
+\frac{\beta^2}{\gamma^2}}.
\end{equation}
A similar analysis to that done in \cite{Aharony:2005hm} shows us
the spectrum must be
\begin{equation}
|\omega|=n+\frac{1}{2}+s
\end{equation}
for $n$ a non-negative integer.  Now, we would like to see if we can
match $s$ to the set of half integers $q/2$.  For non-compact $y$,
we find a continuous spectrum; for compact $y$, presuming $y \in
[0,2\pi]$, we find $\beta=q/2$, where $q$ ranges over the integers.
One can quickly check that only $\alpha=0$ and $\gamma=2$ will allow
$s$ to range over the set of half integers given by $q/2$.  Thus,
only the direct product spacetime with equal characteristic size for
$AdS_2$ and $S^1$ will produce the desired spectrum.

\bibliographystyle{JHEP}
\bibliography{alpha2}

\providecommand{\href}[2]{#2}\begingroup\raggedright\begin{thebibliography}{10}

\bibitem{Nakayama:2004vk}
Y.~Nakayama, {\it Liouville field theory: A decade after the revolution},  {\em
  Int. J. Mod. Phys.} {\bf A19} (2004) 2771--2930,
  [\href{http://xxx.lanl.gov/abs/hep-th/0402009}{{\tt hep-th/0402009}}].

\bibitem{Ginsparg:1993is}
P.~H. Ginsparg and G.~W. Moore, {\it Lectures on 2-d gravity and 2-d string
  theory},  \href{http://xxx.lanl.gov/abs/hep-th/9304011}{{\tt
  hep-th/9304011}}.

\bibitem{Klebanov:1991qa}
I.~R. Klebanov, {\it String theory in two-dimensions},
  \href{http://xxx.lanl.gov/abs/hep-th/9108019}{{\tt hep-th/9108019}}.

\bibitem{Alexandrov:2003ut}
S.~Alexandrov, {\it Matrix quantum mechanics and two-dimensional string theory
  in non-trivial backgrounds},
  \href{http://xxx.lanl.gov/abs/hep-th/0311273}{{\tt hep-th/0311273}}.

\bibitem{Martinec:2004td}
E.~J. Martinec, {\it Matrix models and 2d string theory},
  \href{http://xxx.lanl.gov/abs/hep-th/0410136}{{\tt hep-th/0410136}}.

\bibitem{McGreevy:2003kb}
J.~McGreevy and H.~L. Verlinde, {\it Strings from tachyons: The c = 1 matrix
  reloaded},  {\em JHEP} {\bf 12} (2003) 054,
  [\href{http://xxx.lanl.gov/abs/hep-th/0304224}{{\tt hep-th/0304224}}].

\bibitem{Douglas:2003up}
M.~R. Douglas {\em et~al.}, {\it A new hat for the $c = 1$ matrix model},
  \href{http://xxx.lanl.gov/abs/hep-th/0307195}{{\tt hep-th/0307195}}.

\bibitem{Takayanagi:2003sm}
T.~Takayanagi and N.~Toumbas, {\it A matrix model dual of type 0B string theory
  in two dimensions},  {\em JHEP} {\bf 07} (2003) 064,
  [\href{http://xxx.lanl.gov/abs/hep-th/0307083}{{\tt hep-th/0307083}}].

\bibitem{Horava:2005tt}
P.~Ho\v{r}ava and C.~A. Keeler, {\it Noncritical M-theory in 2+1 dimensions as
  a nonrelativistic Fermi liquid},
  \href{http://xxx.lanl.gov/abs/hep-th/0508024}{{\tt hep-th/0508024}}.

\bibitem{Horava:2005wm}
P.~Ho\v{r}ava and C.~A. Keeler, {\it Thermodynamics of noncritical M-theory and
  the topological A-model},  {\em Nucl. Phys.} {\bf B745} (2006) 1--28,
  [\href{http://xxx.lanl.gov/abs/hep-th/0512325}{{\tt hep-th/0512325}}].

\bibitem{Maldacena:2005he}
J.~M. Maldacena and N.~Seiberg, {\it Flux-vacua in two dimensional string
  theory},  {\em JHEP} {\bf 09} (2005) 077,
  [\href{http://xxx.lanl.gov/abs/hep-th/0506141}{{\tt hep-th/0506141}}].

\bibitem{Strominger:2003tm}
A.~Strominger, {\it A matrix model for AdS(2)},  {\em JHEP} {\bf 03} (2004)
  066, [\href{http://xxx.lanl.gov/abs/hep-th/0312194}{{\tt hep-th/0312194}}].

\bibitem{Ho:2004qp}
P.-M. Ho, {\it Isometry of AdS(2) and the c = 1 matrix model},  {\em JHEP} {\bf
  05} (2004) 008, [\href{http://xxx.lanl.gov/abs/hep-th/0401167}{{\tt
  hep-th/0401167}}].

\bibitem{Aharony:2005hm}
O.~Aharony and A.~Patir, {\it The conformal limit of the 0A matrix model and
  string theory on AdS(2)},  {\em JHEP} {\bf 11} (2005) 052,
  [\href{http://xxx.lanl.gov/abs/hep-th/0509221}{{\tt hep-th/0509221}}].

\bibitem{Gross:1987kz}
D.~J. Gross and P.~F. Mende, {\it The high-energy behavior of string scattering
  amplitudes},  {\em Phys. Lett.} {\bf B197} (1987) 129.

\bibitem{Gross:1987ar}
D.~J. Gross and P.~F. Mende, {\it String theory beyond the Planck scale},  {\em
  Nucl. Phys.} {\bf B303} (1988) 407.

\bibitem{Witten:1988zd}
E.~Witten, {\it Space-time and topological orbifolds},  {\em Phys. Rev. Lett.}
  {\bf 61} (1988) 670.

\bibitem{Amati:1987uf}
D.~Amati, M.~Ciafaloni, and G.~Veneziano, {\it Classical and quantum gravity
  effects from planckian energy superstring collisions},  {\em Int. J. Mod.
  Phys.} {\bf A3} (1988) 1615--1661.

\bibitem{Amati:1987wq}
D.~Amati, M.~Ciafaloni, and G.~Veneziano, {\it Superstring collisions at
  planckian energies},  {\em Phys. Lett.} {\bf B197} (1987) 81.

\bibitem{Amati:1988tn}
D.~Amati, M.~Ciafaloni, and G.~Veneziano, {\it Can space-time be probed below
  the string size?},  {\em Phys. Lett.} {\bf B216} (1989) 41.

\bibitem{Gross:1988ue}
D.~J. Gross, {\it High-energy symmetries of string theory},  {\em Phys. Rev.
  Lett.} {\bf 60} (1988) 1229.

\bibitem{Atick:1988si}
J.~J. Atick and E.~Witten, {\it The Hagedorn transition and the number of
  degrees of freedom of string theory},  {\em Nucl. Phys.} {\bf B310} (1988)
  291--334.

\bibitem{Dijkgraaf:2004te}
R.~Dijkgraaf, S.~Gukov, A.~Neitzke, and C.~Vafa, {\it Topological M-theory as
  unification of form theories of gravity},  {\em Adv. Theor. Math. Phys.} {\bf
  9} (2005) 603--665, [\href{http://xxx.lanl.gov/abs/hep-th/0411073}{{\tt
  hep-th/0411073}}].

\bibitem{Vasiliev:1992gr}
M.~A. Vasiliev, {\it Unfolded representation for relativistic equations in
  (2+1) anti-de Sitter space},  {\em Class. Quant. Grav.} {\bf 11} (1994)
  649--664.

\bibitem{Vasiliev:1992ix}
M.~A. Vasiliev, {\it Equations of motion for d = 3 massless fields interacting
  through Chern-Simons higher spin gauge fields},  {\em Mod. Phys. Lett.} {\bf
  A7} (1992) 3689--3702.

\bibitem{Shaynkman:2001ip}
O.~V. Shaynkman and M.~A. Vasiliev, {\it Higher spin conformal symmetry for
  matter fields in 2+1 dimensions},  {\em Theor. Math. Phys.} {\bf 128} (2001)
  1155--1168, [\href{http://xxx.lanl.gov/abs/hep-th/0103208}{{\tt
  hep-th/0103208}}].

\bibitem{Bekaert:2005vh}
X.~Bekaert, S.~Cnockaert, C.~Iazeolla, and M.~A. Vasiliev, {\it Nonlinear
  higher spin theories in various dimensions},
  \href{http://xxx.lanl.gov/abs/hep-th/0503128}{{\tt hep-th/0503128}}.

\bibitem{Gukov:2003yp}
S.~Gukov, T.~Takayanagi, and N.~Toumbas, {\it Flux backgrounds in 2d string
  theory},  {\em JHEP} {\bf 03} (2004) 017,
  [\href{http://xxx.lanl.gov/abs/hep-th/0312208}{{\tt hep-th/0312208}}].

\bibitem{Davis:2004xb}
J.~L. Davis, L.~A. Pando~Zayas, and D.~Vaman, {\it On black hole thermodynamics
  of 2-d type 0A},  {\em JHEP} {\bf 03} (2004) 007,
  [\href{http://xxx.lanl.gov/abs/hep-th/0402152}{{\tt hep-th/0402152}}].

\bibitem{Danielsson:2004xf}
U.~H. Danielsson, J.~P. Gregory, M.~E. Olsson, P.~Rajan, and M.~Vonk, {\it Type
  0A 2d black hole thermodynamics and the deformed matrix model},  {\em JHEP}
  {\bf 04} (2004) 065, [\href{http://xxx.lanl.gov/abs/hep-th/0402192}{{\tt
  hep-th/0402192}}].

\bibitem{deAlfaro:1976je}
V.~de~Alfaro, S.~Fubini, and G.~Furlan, {\it Conformal invariance in quantum
  mechanics},  {\em Nuovo Cim.} {\bf A34} (1976) 569.

\bibitem{Aharony:2006hz}
O.~Aharony, A.~B. Clark, and A.~Karch, {\it The CFT/AdS correspondence, massive
  gravitons and a connectivity index conjecture},  {\em Phys. Rev.} {\bf D74}
  (2006) 086006, [\href{http://xxx.lanl.gov/abs/hep-th/0608089}{{\tt
  hep-th/0608089}}].

\bibitem{Kiritsis:2006hy}
E.~Kiritsis, {\it Product CFTs, gravitational cloning, massive gravitons and
  the space of gravitational duals},  {\em JHEP} {\bf 11} (2006) 049,
  [\href{http://xxx.lanl.gov/abs/hep-th/0608088}{{\tt hep-th/0608088}}].

\bibitem{Klebanov:2004ya}
I.~R. Klebanov and J.~M. Maldacena, {\it Superconformal gauge theories and
  non-critical superstrings},  {\em Int. J. Mod. Phys.} {\bf A19} (2004)
  5003--5016, [\href{http://xxx.lanl.gov/abs/hep-th/0409133}{{\tt
  hep-th/0409133}}].

\bibitem{Horne:1988jf}
J.~H. Horne and E.~Witten, {\it Conformal gravity in three-dimensions as a
  gauge theory},  {\em Phys. Rev. Lett.} {\bf 62} (1989) 501--504.

\bibitem{Witten:1988hc}
E.~Witten, {\it (2+1)-dimensional gravity as an exactly soluble system},  {\em
  Nucl. Phys.} {\bf B311} (1988) 46.

\bibitem{Fradkin:1987ks}
E.~S. Fradkin and M.~A. Vasiliev, {\it On the gravitational interaction of
  massless higher spin fields},  {\em Phys. Lett.} {\bf B189} (1987) 89--95.

\bibitem{Blencowe:1988gj}
M.~P. Blencowe, {\it A consistent interacting massless higher spin field theory
  in d = (2+1)},  {\em Class. Quant. Grav.} {\bf 6} (1989) 443.

\bibitem{Pope:1989vj}
C.~N. Pope and P.~K. Townsend, {\it Conformal higher spin in (2+1)-dimensions},
   {\em Phys. Lett.} {\bf B225} (1989) 245.

\bibitem{Fradkin:1989xt}
E.~S. Fradkin and V.~Y. Linetsky, {\it A superconformal theory of massless
  higher spin fields in d = (2+1)},  {\em Mod. Phys. Lett.} {\bf A4} (1989)
  731.

\bibitem{Didenko:2006zd}
V.~E. Didenko, A.~S. Matveev, and M.~A. Vasiliev, {\it Btz black hole as
  solution of 3d higher spin gauge theory},
  \href{http://xxx.lanl.gov/abs/hep-th/0612161}{{\tt hep-th/0612161}}.

\bibitem{Vasiliev:1986qx}
M.~A. Vasiliev, {\it Extended higher spin superalgebras and their realizations
  in terms of quantum operators},  {\em Fortsch. Phys.} {\bf 36} (1988) 33--62.

\bibitem{Prokushkin:1999gc}
S.~F. Prokushkin, A.~Y. Segal, and M.~A. Vasiliev, {\it Coordinate-free action
  for AdS(3) higher-spin-matter systems},  {\em Phys. Lett.} {\bf B478} (2000)
  333--342, [\href{http://xxx.lanl.gov/abs/hep-th/9912280}{{\tt
  hep-th/9912280}}].

\bibitem{McGreevy:2003dn}
J.~McGreevy, S.~Murthy, and H.~L. Verlinde, {\it Two-dimensional superstrings
  and the supersymmetric matrix model},  {\em JHEP} {\bf 04} (2004) 015,
  [\href{http://xxx.lanl.gov/abs/hep-th/0308105}{{\tt hep-th/0308105}}].

\bibitem{Verlinde:2004gt}
H.~L. Verlinde, {\it Superstrings on AdS(2) and superconformal matrix quantum
  mechanics},  \href{http://xxx.lanl.gov/abs/hep-th/0403024}{{\tt
  hep-th/0403024}}.

\bibitem{Takayanagi:2004ge}
T.~Takayanagi, {\it Comments on 2d type IIA string and matrix model},  {\em
  JHEP} {\bf 11} (2004) 030,
  [\href{http://xxx.lanl.gov/abs/hep-th/0408086}{{\tt hep-th/0408086}}].

\bibitem{Seiberg:2005bx}
N.~Seiberg, {\it Observations on the moduli space of two dimensional string
  theory},  {\em JHEP} {\bf 03} (2005) 010,
  [\href{http://xxx.lanl.gov/abs/hep-th/0502156}{{\tt hep-th/0502156}}].

\bibitem{Sundborg:2000wp}
B.~Sundborg, {\it Stringy gravity, interacting tensionless strings and massless
  higher spins},  {\em Nucl. Phys. Proc. Suppl.} {\bf 102} (2001) 113--119,
  [\href{http://xxx.lanl.gov/abs/hep-th/0103247}{{\tt hep-th/0103247}}].

\bibitem{Witten:2001js}
E.~Witten, {\it Spacetime reconstruction},  {\em talk at JHS60} (November 2001)
  http://www.theory.caltech.edu/jhs60/witten/1.html.

\bibitem{Sezgin:2002rt}
E.~Sezgin and P.~Sundell, {\it Massless higher spins and holography},  {\em
  Nucl. Phys.} {\bf B644} (2002) 303--370,
  [\href{http://xxx.lanl.gov/abs/hep-th/0205131}{{\tt hep-th/0205131}}].

\bibitem{Klebanov:2002ja}
I.~R. Klebanov and A.~M. Polyakov, {\it AdS dual of the critical O(N) vector
  model},  {\em Phys. Lett.} {\bf B550} (2002) 213--219,
  [\href{http://xxx.lanl.gov/abs/hep-th/0210114}{{\tt hep-th/0210114}}].

\bibitem{Girardello:2002pp}
L.~Girardello, M.~Porrati, and A.~Zaffaroni, {\it 3-d interacting CFTs and
  generalized Higgs phenomenon in higher spin theories on AdS},  {\em Phys.
  Lett.} {\bf B561} (2003) 289--293,
  [\href{http://xxx.lanl.gov/abs/hep-th/0212181}{{\tt hep-th/0212181}}].

\bibitem{Sagnotti:2003qa}
A.~Sagnotti and M.~Tsulaia, {\it On higher spins and the tensionless limit of
  string theory},  {\em Nucl. Phys.} {\bf B682} (2004) 83--116,
  [\href{http://xxx.lanl.gov/abs/hep-th/0311257}{{\tt hep-th/0311257}}].

\bibitem{Petkou:2004nu}
A.~C. Petkou, {\it Holography, duality and higher-spin theories},
  \href{http://xxx.lanl.gov/abs/hep-th/0410116}{{\tt hep-th/0410116}}.

\bibitem{Sagnotti:2005ns}
A.~Sagnotti, E.~Sezgin, and P.~Sundell, {\it On higher spins with a strong
  Sp(2,R) condition},  \href{http://xxx.lanl.gov/abs/hep-th/0501156}{{\tt
  hep-th/0501156}}.

\bibitem{Francia:2006hp}
D.~Francia and A.~Sagnotti, {\it Higher-spin geometry and string theory},  {\em
  J. Phys. Conf. Ser.} {\bf 33} (2006) 57,
  [\href{http://xxx.lanl.gov/abs/hep-th/0601199}{{\tt hep-th/0601199}}].

\bibitem{Horava:1997dd}
P.~Ho\v{r}ava, {\it M-theory as a holographic field theory},  {\em Phys. Rev.}
  {\bf D59} (1999) 046004, [\href{http://xxx.lanl.gov/abs/hep-th/9712130}{{\tt
  hep-th/9712130}}].

\end{thebibliography}\endgroup
\end{document}